\documentclass[12pt]{article}
\usepackage{color}
\usepackage{amssymb}
\usepackage{amsmath}
\usepackage[dvips]{graphicx}
\usepackage{mathrsfs}
\usepackage[mathscr]{eucal}
\usepackage{caption}
\usepackage{subcaption}
\newcommand{\bs}{\boldsymbol}
\author{Mariusz Adamski$^1$, Janusz J{\c{e}}drzejewski$^1$ and Taras Krokhmalskii$^2$\\
$^1$Institute of Theoretical Physics, University of Wroc\l aw,\\
pl. Maksa Borna 9, 50--204 Wroc\l aw, Poland\\
$^2$Institute for Condensed Matter Physics,\\
1 Svientsitski Street, 79011 Lviv, Ukraine}
\title{Quantum critical scaling of fidelity in BCS-like model}

\begin{document}
\maketitle

\begin{abstract}
\noindent
We study scaling of the ground-state fidelity in neighborhoods of quantum critical points in a model of interacting spinful  fermions -- a BCS-like model. Due to the exact diagonalizability of the model, in one and higher dimensions, scaling  of the ground-state fidelity can be analyzed numerically with great accuracy, not only for small systems but also for macroscopic ones, together with the crossover region between them. Additionally, in one-dimensional case we have been able to derive a number  of analytical formulae for fidelity and show that they accurately fit our numerical results; these results are reported in the article. Besides regular critical points and their neighborhoods, where well-known scaling laws are obeyed, there is the multi-critical point and critical points in its proximity where anomalous scaling behavior is found. We consider also scaling of fidelity in neighborhoods of critical points where fidelity oscillates strongly as the system size or the chemical potential is varied.  Our results for a one-dimensional version of a BCS-like model are compared with those obtained by Rams and Damski \cite{rams PRA 11} in similar studies of a quantum spin chain -- an anisotropic XY model in transverse magnetic field.
\end{abstract}

\section{\label{intro} Introduction}

In the last decade quantum phase transitions and quantum critical phenomena continue to be a subject of great interest, vigorously studied in condensed matter physics. Both, experimental and theoretical developments point  out to the crucial role that quantum phase transitions play in physics of high-$T_c$ superconductors, rare-earth magnetic systems, heavy-fermion systems or two-dimensional electrons liquids exibiting fractional quantum Hall effect \cite{sachdev QPTs}, \cite{vojta QPTs}. The so called classical, thermal phase transitions originate from thermal fluctuations and are mathematically manifested as singularities in temperature and other thermodynamic parameters of various thermodynamic functions, and such correlation quantities like the correlation length, at nonzero temperatures. In contrast, quantum phase transitions originate from purely quantum fluctuations and are mathematically manifested as singularities in system parameters of the ground-state energy density, which is also  the zero-temperature limit of the internal energy density. Naturally, singularities of thermodynamic functions appear only in the thermodynamic limit. The importance of quantum phase transitions for physics and the related wide interest in such  transitions stems from the fact that, while a quantum phase transition is exhibited by ground states, hence often termed a zero-temperature phenomenon, its existence in a system exerts a great impact on the behavior of that system also at nonzero temperatures.

Typically, theoretical studies of quantum phase transitions can be conducted much along the same lines as in the case of thermal phase transitions: one considers possibly simple models, studies the eigenvalue problem of  Hamiltonians,  excitation gaps, constructs local order parameters, determines broken symmetries, calculates the correlation functions. However in distinction to thermal phase transitions, completely new approaches to the subject have been put forward, which are based on ideas of the quantum-information science. The central object of quantum-information approaches to many-body systems is the ground state; there is no need to consider order parameters and broken symmetries, which makes these approaches particularly useful for quantum phase transitions, which are not characterized in such terms, like topological phase transitions. One of the quantum-information approaches, the entanglement approach, with longer than a decade history, exploits the notion of entanglement of quantum states in the context of its applications to many-body systems, for review see \cite{vedral 08}; this approach is out of the scope of our article. Another approach, the fidelity approach, put forward by Zanardi and Paunkovi\'c \cite{zanardi paunkovic 06}, is a relatively new idea of using the quantum fidelity of ground states as a probe of quantum phase transitions. A comprehensive exposition of the fidelity approach can be found in the review by Gu \cite{gu 10}.

In what follows we shall study continuous quantum phase transitions, and it is the fidelity approach that will be used for that purpose.
Let ${\bs \lambda}$ be a vector whose components are those parameters of the considered system's Hamiltonian that drive a quantum phase transition, and ${\bs e}$ - a unit vector in the space of those parameters. Then, on varying parameter $\delta$ the vectors ${\bs \lambda} + \delta {\bs e}$ scan a neighborhood of ${\bs \lambda}$ along direction $\bs e$. For given ${\bs \lambda}$ and $\delta$, the ground-state fidelity at ${\bs \lambda}$ in direction $\bs e$,
${\mathscr{F}}_{\bs e}({\bs\lambda}, \delta)$, is the absolute value of the overlap of the ground states $|{\bs \lambda} \pm \delta {\bs e} \rangle$  at the points ${\bs \lambda} \pm \delta {\bs e}$,
\begin{equation}
{\mathscr{F}}_{\bs e}({\bs\lambda}, \delta) =
|\langle {\bs\lambda} - \delta {\bs e}|{\bs\lambda} + \delta {\bs e} \rangle|.
\label{fidelity}
\end{equation}
A list of general, system independent, properties of ${\mathscr{F}}_{\bs e}({\bs\lambda}, \delta)$ can be found in \cite{gu 10}.
The transition point of a continuous quantum phase transition, i.e. the quantum critical point,
denoted ${\bs \lambda}_c$, is characterized by the power-law divergence of the correlation length $\xi({\bs\lambda})$, as the quantum critical point is approached:
$\xi({\bs\lambda}) \sim |{\bs \lambda} - {\bs \lambda}_c|^{-\nu}$. Alternatively, ${\bs \lambda}_c$ can be defined as the point where the gap between the ground-state energy and the energy of the lowest excited state vanishes. In reference \cite{zanardi paunkovic 06} it was demonstrated that the quantum critical point ${\bs \lambda}_c$ can be identified as the minimum of fidelity, as ${\bs \lambda}$ is varied. However, the fidelity approach seeks an answer to a more general question: does the behavior of quantum fidelity in a neighborhood of a quantum critical point encode not only the location of that point but also some universal information about the underlying quantum phase transition?
First results, pointing towards a positive answer to the raised question by providing some finite-size critical scaling of geometric tensors, have been obtained by Venuti and Zanardi \cite{venuti zanardi 07}.

According to finite-size scaling theories, the properties of a system are close to those at the thermodynamic limit, we say the system is macroscopic, if the linear size of the system, $L$, is much greater than $\xi({\bs\lambda})$. In the opposite limit we observe the so called "small-system" properties. Concerning finite-size scaling properties of fidelity ${\mathscr{F}}_{\bs e}({\bs\lambda}, \delta)$, it is expected that the characteristic length of the system, that differentiates between small and macroscopic system, is given by
${\tilde{\xi}}_{\bs e}({\bs\lambda}, \delta)$, which is the smaller of the two correlation lengths
$\xi({\bs \lambda} \pm \delta {\bs e})$. In other words, the crossover between small-system and  macroscopic-system properties occurs, when the effective linear size of a system,
$L/{\tilde{\xi}}_{\bs e}({\bs\lambda}, \delta)$, satisfies the crossover condition:
\begin{equation}
L/{\tilde{\xi}}_{\bs e}({\bs\lambda}, \delta) \sim 1 .
\label{crossover}
\end{equation}
There are numerous papers devoted to critical scaling of small-system fidelity,  see \cite{gu 10},
\cite{venuti zanardi 07}, \cite{albuquerque 10}, and references quoted there.
Typically,  whether ${\bs\lambda}$ is close to or away from ${\bs \lambda}_c$, small-system fidelity can be Taylor-expanded in $\delta$,
\begin{equation}
{\mathscr{F}}_{\bs e}({\bs\lambda}, \delta) =
1 - \frac{ \delta^2}{2} \chi_{\bs e}({\bs\lambda}) + \ldots ,
\label{susceptibility}
\end{equation}
where the first order term vanishes because of the symmetry of fidelity in $\delta$ at zero. The  coefficient of the second order term, $\chi_{\bs e}({\bs\lambda})$, is known as the fidelity susceptibility. One expects some universal scaling properties of $\chi_{\bs e}({\bs\lambda})$, provided $\bs\lambda$ is sufficiently close to a quantum critical point ${\bs\lambda}_c$, where the correlation  length diverges:
$\xi({\bs \lambda}_c \pm \delta {\bs e})\sim |\delta|^{-\nu}$.
Fairly general, model-independent, arguments provide us with finite-size scaling of the fidelity susceptibility at ${\bs\lambda}_c$  \cite{albuquerque 10}:
\begin{equation}
\chi_{\bs e}({\bs\lambda}_c) \sim L^{2/\nu},
\label{susc scaling 1}
\end{equation}
that is, in the small-system regime
\begin{equation}
-\ln {\mathscr{F}}_{\bs e}({\bs\lambda}_c, \delta) \sim \delta^2 L^{2/\nu}.
\label{susc scaling 2}
\end{equation}
In the macroscopic-system regime, quantum phase transitions have been studied by means of the so called fidelity per site, a quantity whose logarithm is equal to $N^{-1} \ln {\mathscr{F}}_{\bs e}({\bs\lambda}, \delta)$, where $N=L^d$ is the number of sites in a $d$-dimensional system \cite{barjaktarevic 08}, \cite{zhou 08}.
In this regime expansion (\ref{susceptibility}) is not valid near critical points
\cite{barjaktarevic 08}-\cite{rams PRA 11}.
However, critical scaling of a macroscopic-system fidelity has been considered only very recently by Rams and Damski \cite{rams PRL 11},\cite{rams PRA 11}. These authors found that, while for small-systems the fidelity scaling is totaly insensitive to the way the critical point ${\bs\lambda}_c$ is approached  by the points ${\bs \lambda} \pm \delta {\bs e}$ (i.e. for instance, whether they are located on one side of the critical point or on the opposite sides), in the case of macroscopic-system the way of approaching the critical point matters.  To make this explicit, Rams and Damski substituted ${\bs\lambda}_c + c \delta {\bs e}$ for ${\bs\lambda}$. By choosing the value of the parameter  $c$, the above mentioned location of  the two points
${\bs\lambda}_c +  c \delta {\bs e} \pm \delta {\bs e}$ with respect to the critical point can be controlled. If $|c|>1$ or  $|c|<1$, then both points are located on one side or on opposite sides of ${\bs\lambda}_c$, respectively. If $|c|=1$, then one of the points coincides with ${\bs\lambda}_c$. Now, the above mentioned independence of the small-system-fidelity scaling on the way the critical point ${\bs\lambda}_c$ is approached  by the points
${\bs \lambda} \pm \delta {\bs e}$ can be expressed as follows:
\begin{equation}
-\ln {\mathscr{F}}_{\bs e}({\bs\lambda}_c + c \delta {\bs e}, \delta) \sim \delta^2 L^{2/\nu}.
\label{susc scaling 3}
\end{equation}
In contrast to the small-system case, the fidelity scaling law for macroscopic systems, derived by Rams and Damski \cite{rams PRL 11}, makes the dependence on parameter $c$ explicit, and reads
\begin{equation}
- \ln {\mathscr{F}}_{\bs e}({\bs\lambda}_c + c \delta {\bs e}, \delta) \sim |\delta|^{d\nu} N {\mathscr{A}}_{\bs e}(c),
\label{fid scal}
\end{equation}
where ${\mathscr{A}}_{\bs e}(c)$ is the scaling function.
 Scaling law (\ref{susc scaling 3}) is expected to hold provided the thermodynamic limit of
$N^{-1}\ln {\mathscr{F}}_{\bs e}({\bs\lambda}, \delta)$ does exist, there is only one characteristic length scale given by $\xi$, and $d \nu <2$ \cite{rams PRL 11},\cite{rams PRA 11}.

The testing ground for new developments in theory of quantum phase transitions consists mainly of two paradigmatic quantum spin chains, an Ising chain and an (isotropic or anisotropic)  XY chain, both in a transverse magnetic field. In this respect, the fidelity approach is no exception as can be seen in references \cite{gu 10} -
%\cite{barjaktarevic 08}, \cite{zhou 08}, \cite{rams PRL 11},
\cite{rams PRA 11}.
This stems from the exact diagonalizability of both models, which makes possible some analytical analysis. In those models where such an analysis is not feasible, only rather small systems are studied numerically, with no perspective to reach a macroscopic regime \cite{gu 10}. Due to Jordan-Wigner transformation the two mentioned quantum spin chains are equivalent to one-dimensional spinless-fermion lattice gases that are exactly diagonalizable for a specific boundary condition. Beyond one dimension those models are not exactly diagonalizable, hence less useful for testing theories of quantum phase transitions.

To test the fidelity approach, in particular critical scaling of fidelity, from small to macroscopic systems, we consider spinful fermion models that are exactly diagonalizable in any dimension, for periodic, antiperiodic or twisted boundary conditions. These models originate from the two-dimensional model of d-wave superconductivity proposed by Sachdev \cite{sachdev 02}, \cite{sachdev QPTs}. Naturally, the restriction to one-dimension is of special merit, since besides accurate numerical results, we have been able to derive analytical results for two-point correlations functions, the correlation length, and fidelities in small- and macroscopic-system regimes. These results constitute the contents of our paper. Results for two-dimensional systems will be presented elsewhere.

The paper is organized as follows. In section \ref{model} we present our model and basic facts relevant for subsequent discussion. Then, in 6 subsections of section \ref{criticality}, we study critical scaling of fidelity, in small- and macroscopic-system regimes, and the crossover region between them, in vicinities of different kinds of quantum critical points of the model. The obtained results are summarized in section \ref{summ}.

\section{\label{model} The model, its quantum critical points, ground-state two-point correlation function and fidelity}

We consider a $d$-dimensional spinful fermion model, given by the Hamiltonian,
\begin{equation}
H =- \frac{t}{2}  \sum_{{\bs l},i,\sigma}  \left(
a^{\dagger}_{{\bs l},\sigma} a_{{\bs l}+{\bs e}_i,\sigma}
+ \mathrm{h.c.} - \mu a^{\dagger}_{{\bs l},\sigma} a_{{\bs l},\sigma} \right)
- \frac{1}{2} \sum_{{\bs l},i} J_i  \left(
a^{\dagger}_{{\bs l},\uparrow}a^{\dagger}_{{\bs l}+{\bs e}_i,\downarrow} -
a^{\dagger}_{{\bs l},\downarrow}a^{\dagger}_{{\bs l}+{\bs e}_i,\uparrow} \right)
+ \mathrm{h.c},
\label{ham1}
\end{equation}
where $a^{\dagger}_{{\bs l},\sigma}$, $a_{{\bs l},\sigma}$ stand for creation and annihilation operators of an electron with spin projection $\sigma = \uparrow,\downarrow $ in a state localized at site ${\bs l}$ of a $d$-dimensional hypercubic lattice  with some boundary conditions, respectively, and ${\bs e}_i$, with $i=1,\ldots,d$ are unit vectors whose $m$-th component is $\delta_{i,m}$. Both summations in (\ref{ham1}) are over pairs of nearest neighbors, with each pair counted once. The real and positive parameter $t$ is the nearest-neighbor hoping intensity, $\mu$ -- the chemical potential, and $J_i$ -- the coupling constant in direction ${\bs e}_i$ of the gauge-symmetry breaking interaction, in general complex. Naturally, we can express the parameters $\mu$ and $J_i$ in units of $t$, while the lengths of the underlying lattice in units of the lattice constant, preserving the original notation.
We emphasize that in distinction to \cite{sachdev 02}, \cite{sachdev QPTs}, where Hamiltonian (\ref{ham1}) was derived, here the parameters $\mu$ and $J_i$ are independent.
Moreover, we impose on the system the antiperiodic boundary conditions, for which the grid of wave vectors
${\bs k}$ (quasimomenta) is given as follows: a component $k_i$ of ${\bs k}$ assumes the values $k_i=(2n+1-L)\pi /L$, with $n=0,1, \ldots ,L-1$, where  $L$ is the size of the underlying lattice in direction ${\bs e}_i$, and we choose $L$ divisible by 4. The values of $k_i$ are distributed symmetrically about zero, the closest to zero points are $\pm \pi/L$ and the closest to $\pm \pi/2$ points are $\pm \pi/2 \pm \pi/L$.
Then, after passing from the site-localized to the plane-wave basis,  Hamiltonian (\ref{ham1}) assumes the form
\begin{equation}
H = \sum_{{\bs k},\sigma}\varepsilon_{\bs k} c^{\dagger}_{{\bs k},\sigma} c_{{\bs k},\sigma}
- \sum_{{\bs k}}  \left( \sum_i J_i \cos{k_i} \right) c^{\dagger}_{{\bs k},\uparrow}c^{\dagger}_{-{\bs k},\downarrow} + \mathrm{h.c.},
\label{ham2}
\end{equation}
where
\begin{equation}
\varepsilon_{\bs k} = \sum_i \cos k_i - \mu.
\label{epsilon}
\end{equation}
Formally, Hamiltonian (\ref{ham2}) differs from the well-known BCS Hamiltonian of s-wave superconductivity by the presence of  $\cos k_i$ factors in the gauge-symmetry breaking term. Such Hamiltonians can readily be diagonalized by means of the Bogoliubov transformation. The dispersion relation of quasi-particles reads
\begin{equation}
E_{\bs k} + \sum_{\bs k} \left( \varepsilon_{\bs k} - E_{\bs k} \right),
\label{spectrum}
\end{equation}
where
\begin{equation}
E_{\bs k} =\sqrt{ \varepsilon_{\bs k}^2 + \left| \sum_i J_i \cos k_i \right|^2}.
\label{Ek}
\end{equation}
The Hamiltonian (\ref{ham1}) preserves parity; therefore without any loss of generality we can restrict the state-space to the subspace of even number of particles (electrons). In this subspace, the state $|0\rangle_{qp}$ of an unspecified (but even) number of electrons, defined by
\begin{equation}
|0\rangle_{qp}= \prod_{\bs k}(u_{\bs k} + v_{\bs k} c^{\dagger}_{{\bs k},\uparrow}c^{\dagger}_{-{\bs k},\downarrow})|0\rangle,
\label{gs}
\end{equation}
where $|0\rangle$ is the electron vacuum, with $u_{\bs k}$ real and positive,
\begin{equation}
u_{\bs k}= \sqrt{\frac{1}{2}\left(1+ \frac{\varepsilon_{\bs k}}{E_{\bs k}}\right)},
\label{uk}
\end{equation}
and, in general, complex $v_{\bs k}$,
\begin{equation}
|v_{\bs k}|= \sqrt{\frac{1}{2}\left(1- \frac{\varepsilon_{\bs k}}{E_{\bs k}}\right)}, \,\,\,
\arg v_{\bs k} = \arg  \sum_i J_i \cos k_i,
\label{vk}
\end{equation}
is the eigenstate of (\ref{ham2}) to the lowest eigenenergy, $\sum_{\bs k} \left(\varepsilon_{\bs k} - E_{\bs k}  \right)$, i.e. the ground state. This state is the vacuum of elementary excitations (quasi-particles), whose energies, relative to the ground-state energy, are $E_{\bs k}$. As long as our system is finite, the energies $E_{\bs k}$ are the excitation gaps, since $E_{\bs k} > 0$ for all values of ${\bs k}$. However, in the thermodynamic limit, $L \to \infty$, the excitation gaps $E_{\bs k}$ in the spectrum  of quasi-particles may close at special values of wave vector ${\bs k}$, denoted ${\bs k}_c$, and at special values of the chemical potential $\mu$ and coupling constants $J_i$.

One can define several two-point correlation functions for the considered system, which however are simply related with each  other. Each one can be used to calculate the correlation length. We choose the two-point function given by the matrix elements of the ground-state one-body reduced density operator,
$ _{qp}\langle 0|a^{\dagger}_{{\bs 0},\sigma} a_{{\bs r},\sigma}  |0 \rangle_{qp} $. Explicitly,
\begin{equation}
 _{qp}\langle 0|a^{\dagger}_{{\bs 0},\sigma} a_{{\bs r},\sigma}  |0 \rangle_{qp} =
-\frac{1}{2L^d} \sum_{\bs k} \frac{\varepsilon_{\bs k}}{E_{\bs k}} \exp i{\bs k}{\bs r}.
\label{corr_f 1}
\end{equation}
Using the invariance of functions $\varepsilon_{\bs k}$ and $E_{\bs k}$  with respect to reflections in coordinate axes, and then taking the thermodynamic limit
\begin{equation}
\lim_{L \to \infty} { _{qp}\langle } 0|a^{\dagger}_{{\bs 0},\sigma} a_{{\bs r},\sigma}  |0 \rangle_{qp}
\equiv {\mathscr{G}}(r) =
- \frac{1}{2 \pi                                                                                                      ^d} \int_{0 \leq k_j \leq \pi} d{\bs k} \frac{\varepsilon_{\bs k} }{E_{\bs k}}
\prod_{j=1}^d \cos k_j r_j,
\label{corr_f 2}
\end{equation}
where $r=|{\bs r}|$.

Let $|0 \rangle_{qp}$ and $|\tilde{0} \rangle_{qp}$ be two ground states, the first one for parameters $\mu$, $J_i$, and the functions $\varepsilon_{\bs k}$, $E_{\bs k}$, the second one for parameters ${\tilde \mu}$, ${\tilde J_i}$, and the functions ${\tilde \varepsilon_{\bs k}}$, ${\tilde E_{\bs k}}$. As a result of the product structure of the ground states, the quantum fidelity for these states has also a product structure,
\begin{equation}
|_{qp}\langle 0 |\tilde{0} \rangle_{qp}|=
\prod_{\bs k}| \left( u_{\bs k} {\tilde u_{\bs k}} +
|v_{\bs k}| |{\tilde v_{\bs k}}| \exp i (\arg {\tilde v_{\bs k}} - \arg v_{\bs k}) \right) |.
\label{qp_fidelity}
\end{equation}

Further considerations will be restricted to the one-dimensional case, that is in (\ref{ham2}) we set the coupling constants $J_i=0$, for $i>1$, $J_1 \equiv J$ --- a real number, with $J$ independent of $\mu$, and the wave vector ${\bs k}$ reduces to the wave number $k$. Consequently, Hamiltonian (\ref{ham1}) is parameterized by the pairs $(\mu,J)$.
After using (\ref{uk}), (\ref{vk}) the fidelity of two ground states (\ref{gs}) assumes the form
\begin{equation}
|_{qp}\langle 0 |\tilde{0} \rangle_{qp}|=
\prod_{k >0} f(k),\,\, f(k)=
\frac{1}{2}
\left( 1 + \frac{\varepsilon_k {\tilde \varepsilon}_k + J {\tilde J} \cos^2 k}{E_k {\tilde E}_k} \right),
\label{qp_fidelity_1d}
\end{equation}
where the restriction of the product to $k>0$ is a consequence of the symmetry in $k$ at zero of the factors in  product (\ref{qp_fidelity}).

The critical points of the above specified one-dimensional version of (\ref{ham1}) are located at the line $\mu=0$ for $k_c=\pm \pi/2$, and at the line $J=0$, if $-1\leq \mu \leq 1 $, for $k_c=\arccos(\mu)$.
\begin{figure}
\begin{center}
\includegraphics[width=10cm,clip=on]{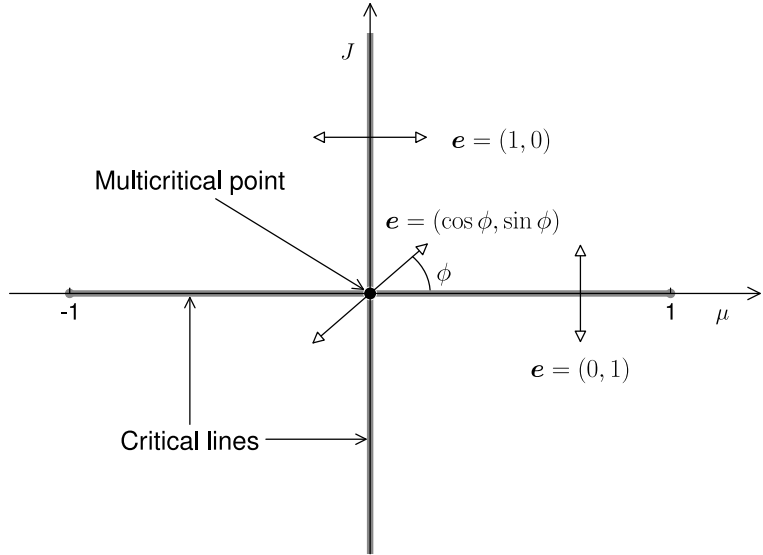}
\caption{\label{diagram} Critical lines in the $(\mu,J)$ plane; the versors $\bs e$ show directions in which neighborhoods of critical points are scanned by the points ${\bs\lambda}_c + (c \pm 1) \delta {\bs e}$.}
\end{center}
\end{figure}
 We note on passing that these lines are symmetry lines of Hamiltonian (\ref{ham1}): at $\mu=0$ line $H$ is hole-particle invariant and at $J=0$ line it is gauge invariant. We do not point out order parameters able to distinguish ground-state phases, since in the fidelity approach to quantum phase transitions they do not play any role (which is one of the advantages of fidelity approach).
Among the mentioned critical points one can distinguish the multicritical point, $\mu=0$ and $J=0$, and two critical end points, $\mu=\pm 1$ and $J=0$, see Fig.\ref{diagram}.

Since for the considerations in the sequel the knowledge of the dependence of the correlation length on $\mu$ and $J$ plays a key role we present a summary of our results obtained in the one-dimensional case \cite{adamski 12}.
In a neighborhood of the line of critical points, $\mu=0$, the large-distance asymptotics of the integral in (\ref{corr_f 2}) has been determined in \cite{adamski 12} by means of coupled analytical and numerical arguments.
It is convenient to introduce an auxiliary function ${\tilde{\mathscr{G}}}(r)$ :
\begin{equation}
{\tilde{\mathscr{G}}}(r)= - \frac{1}{2\pi} \int_0^{\pi} dk \frac{\cos kr}{E_k},
\label{funct g}
\end{equation}
in terms of which
\begin{equation}
{\mathscr{G}}(r) = \frac{1}{2} \left[ {\tilde{\mathscr{G}}}(r+1) + {\tilde{\mathscr{G}}}(r-1) \right]
- \mu {\tilde{\mathscr{G}}}(r).
\label{funct g tilde}
\end{equation}
The large-distance asymptotics of ${\tilde{\mathscr{G}}}(r)$  reads
\begin{equation}
{\tilde{\mathscr{G}}}(r)= \mathscr{C} (-1)^r \cos\left[\left(r+\frac{1}{2}\right)\theta -\frac{\pi}{4}\right] r^{-1/2}e^{-r/\xi},
\label{asymptotics}
\end{equation}
where the distance-independent factor $\mathscr{C}$ is
\begin{eqnarray}
\mathscr{C} = \left( \frac{1}{2\pi^3} \right)^{1/2}(1-\beta^2)^{-1/4} [(1+\mu)^2 + J^2]^{-1/2},
\label{C}
\end{eqnarray}
with
\begin{equation}
\beta = \left(1+\frac{\mu}{1+J^2}\right) \left( \frac{1+J^2}{(1+\mu)^2 + J^2} \right)^{1/2},
\label{beta}
\end{equation}
\begin{equation}
\theta = \arccos \frac{(-\mu)}{1 + J^2},
\label{theta}
\end{equation}
and
\begin{equation}
\xi = \frac{1+J^2}{|J|\mu}.
\label{xi}
\end{equation}
In formulae (\ref{C})-(\ref{xi}), the chemical potential is taken positive, $\mu >0$; they hold for  $\mu /(1+J^2)$ sufficiently small (less than $10^{-2}$ in our calculations).
Apparently, the quantity $\xi$ that follows from formula (\ref{asymptotics}) for the subsidiary function
${\tilde{\mathscr{G}}}(r)$ is the correlation length, as defined by the  large-distance asymptotic behavior of the correlation function ${\mathscr{G}}(r)$, in a neighborhood of critical line $\mu=0$. As $\mu \to 0$ the correlation length diverges with the critical index $\nu=1$.

Unfortunately, we have not been able to find analytic formulae for the correlation length in neighborhoods of critical points located at the line $J=0$ in the interval $-1\leq \mu \leq 1 $. Numerical calculations show that for $0<\mu < 1$, $\xi = \zeta(\mu)|J|^{-1}$, with some function $\zeta(\mu)$, that is again $\nu=1$. However, at the end critical points, $\mu=\pm 1$ and $J=0$, numerical calculations give $\xi \sim |J|^{-1/2}$, i.e. $\nu=1/2$.

Let us adapt the general notation introduced in section \ref{intro} to the considered here model. As the location of critical points is uniquely determined  by pairs $(\mu,J)$, we set ${\bs\lambda} \equiv (\mu,J)$, hence $|{\bs{\lambda}} \rangle \equiv |0\rangle_{qp}$.
Then, in formula (\ref{qp_fidelity_1d}) for fidelity, the functions $\varepsilon_k$, $E_k$, given by (\ref{epsilon}) and (\ref{Ek}), respectively, are calculated at
${\bs\lambda}_c + (c-1) \delta {\bs e}$, while ${\tilde \varepsilon}_k$ and ${\tilde E}_k$ --  at
${\bs\lambda}_c + (c+1) \delta {\bs e}$. Finally, we set
$|_{qp}\langle 0 |\tilde{0} \rangle_{qp}| \equiv{\mathscr{F}}_{\bs e}({\bs{\lambda}}_c, \delta)$.

In what follows, we shall study numerically the sum
\begin{equation}
\sum_{k>0} \left( -\ln f(k) \right) \equiv - \ln {\mathscr{F}}_{\bs e}({\bs{\lambda}}_c, \delta),
\label{fid_sum}
\end{equation}
as a function of parameter $\delta$ for fixed system size $L$, or vice versa, in neighborhoods of various critical points, in small- and macroscopic-system regimes. In all the considered cases the function $\ln f(k)$ is either continuous in the whole interval $[0,\pi]$ or it has an integrable singularity at some $k$ (a discontinuity or a logarithmic divergence). Therefore, in all the considered cases the limit $L \to \infty$ of the Riemann sum corresponding to (\ref{fid_sum}) does exist,
\begin{equation}
\lim_{L \to \infty} - L^{-1} \ln {\mathscr{F}}_{\bs e}({\bs{\lambda}}_c, \delta) =
\frac{1}{2\pi} \int_{0}^{\pi} dk \,\left(- \ln f(k) \right).
\label{fid_integral}
\end{equation}
Consequently, for given sufficiently small $|\delta|$ and sufficiently large $L$
\begin{equation}
- \ln {\mathscr{F}}_{\bs e}({\bs{\lambda}}_c, \delta) \approx
\frac{L}{2\pi} \int_{0}^{\pi} dk \, \left(-\ln f(k) \right)
\label{fid_integral_L}
\end{equation}
approximately, that is in a macroscopic-system regime $- \ln {\mathscr{F}}_{\bs e}({\bs{\lambda}}_c, \delta)$ scales with the system size as $L$.

\section{\label{criticality} Critical scaling of the ground-state fidelity}

Any study of critical behavior involves specifying a critical region, that is a critical point and its neighborhood. The quantum critical points considered in our paper are displayed in Fig. \ref{diagram}. As for neighborhoods, we have chosen line neighborhoods, each one specified by a unit vector ${\bs e}$ and a range of parameter $\delta$, which are scanned by vectors ${\bs\lambda}_c + (c-1) \delta {\bs e}$ and ${\bs\lambda}_c + (c+1) \delta {\bs e}$ on varying $\delta$ (see section \ref{intro}). It might be expected, and it is indeed the case as our studies show, that there is several kinds of specific critical scaling of fidelity, depending on the choice of the critical region. Therefore, our results concerning critical scaling of the ground-state fidelity are presented in six subsections, each one labeled by the specific location of the considered critical points ${\bs \lambda}_c$ and direction ${\bs e}$ of the line neighborhood.

In the sequel, we typically calculate $-\ln {\mathscr{F}}_{\bs e}({\bs{\lambda}}_c, \delta)$ as a function of $\delta$, keeping the linear size $L$ fixed, for a number of $L$;  in fact
$-\ln {\mathscr{F}}_{\bs e}({\bs{\lambda}}_c, \delta)$ depends on $|\delta|$, since
${\mathscr{F}}_{\bs e}({\bs{\lambda}}_c, \delta)= {\mathscr{F}}_{\bs e}({\bs{\lambda}}_c, -\delta)$.
However, sometimes it turned out to be instructive to fix $\delta$ and vary $L$. According to our results concerning the correlation length, summarized in section \ref{model}, in particular formula (\ref{xi}), for a given linear size of the system, we enter the regime of small system by decreasing sufficiently $|\delta|$ or the regime of macroscopic system -- by increasing it sufficiently.
Then, we verify the scaling with $L$ and with $\delta$, and identify  small-system, macroscopic-system and crossover regimes. In many cases we provide analytic approximate formulae for
$-\ln {\mathscr{F}}_{\bs e}({\bs{\lambda}}_c, \delta)$ in small- and macroscopic-system regimes, and compare them with numerical data. In discussing critical scaling properties of fidelity, it is always instructive to study the behavior of the function $f(k)$ of formula (\ref{qp_fidelity_1d}) in neighborhoods of wave numbers $k_c$ for which the excitation gap $E_k$ closes.

Finally, let us note that there is no critical scaling of interest for critical points $|\mu|\leq 1$ at the line $J=0$ with neighborhoods along the $\mu$-axis (${\bs e} =(1,0)$). In this case the considered system is free, its ground state is a Fermi sea and fidelity ${\mathscr{F}}_{(1,0)}((\mu,0), \delta)$ can assume only two values, zero or one. The fidelity vanishes for sufficiently large systems, unless both of the points $(\mu,0) + (c\pm 1) \delta (1,0)$  are located on one side of the $|\mu|\leq 1$ interval.

\subsection{\label{lJ} Across the $\mu {=} 0$ critical line: ${\bs \lambda}_c {=} (0,J)$, $J$ away from $0$, $\bs e {=} (1,0)$}

At critical points ${\bs \lambda}_c = (0,J)$, independently of $J$, the excitation gap $E_k$ closes  at wave number $k_c=\pi/2$. In  line neighborhoods  in direction ${\bs {e}}=(1,0 )$ of those critical points the function $f(k)$ assumes the form
\begin{eqnarray}
f(k)=\frac{1}{2}\left \{ 1+\frac{(\cos k - \mu)(\cos k -\tilde{\mu}) + J^2 \cos^2k}
{\sqrt{ \left[ (\cos k - \mu)^2+J^2\cos^2 k \right] \left[(\cos k - \tilde{\mu})^2+ J^2 \cos^2 k \right]}}
\right \},
\label{f_(0,J)_(1,0)}
\end{eqnarray}
where we set
\begin{eqnarray}
\mu=(c-1)\delta, \quad  \tilde{\mu}=(c+1)\delta,
\end{eqnarray}
with $c>0$ throughout this section.

\begin{figure}
\centering
\begin{subfigure}[b]{0.48\textwidth}
\centering
\includegraphics[width=8cm,clip=on]{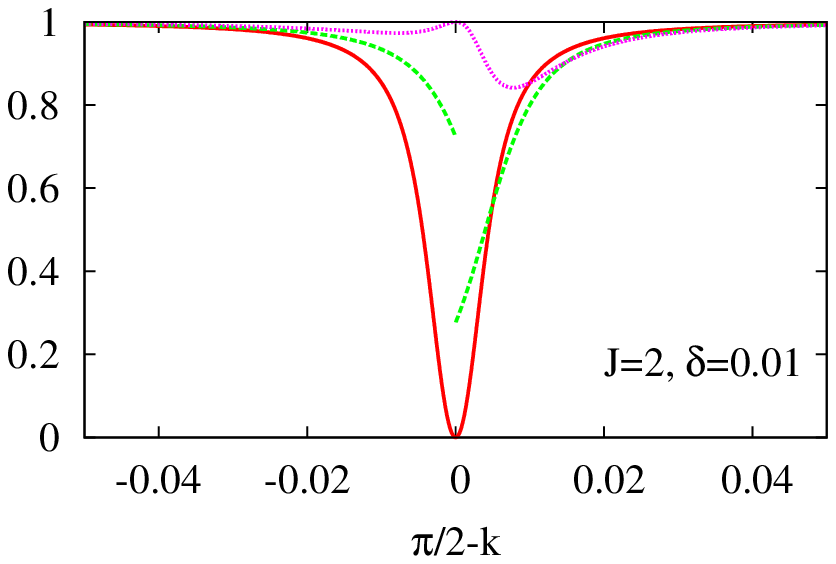}
\end{subfigure}
~
\begin{subfigure}[b]{0.48\textwidth}
\centering
\includegraphics[width=8cm,clip=on]{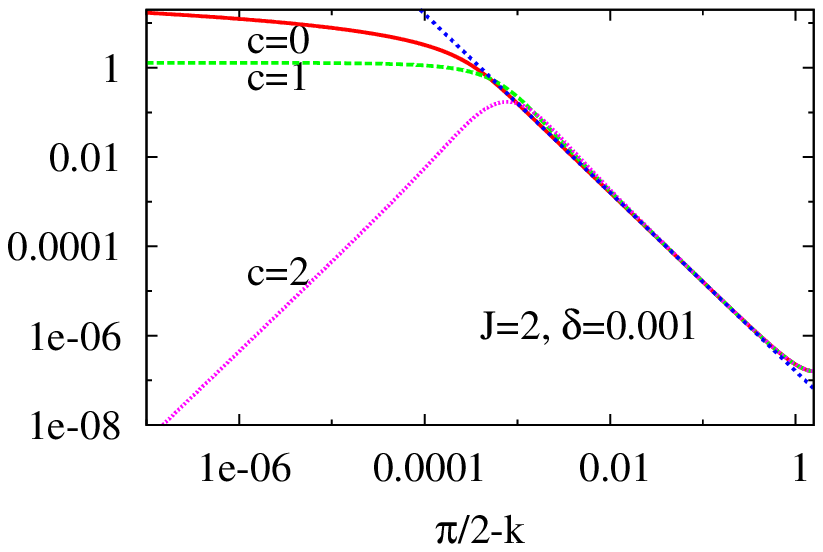}
\end{subfigure}
\caption{\label{f_(0,2)_(1,0)}\label{h_(0,2)_(1,0)}(Color online). The plots of
$f$ (left panel) and $-\ln f$ (right panel) versus $\pi/2 - k$ for three values of $c$:
$c=0$ -- red solid line, $c=1$ -- green dashed line, $c=2$ -- magenta dotted
line. Right panel is in doubly logarithmic scale, and the blue dotted straight
line indicates the $-\ln{\mathscr{F}}_{(1,0)}((0,J), \delta ) \sim \kappa^{-2}$
behavior, and is given by formula (\ref{f_(0,J)_(1,0)_ss}).}
\end{figure}

In this section we consider $J$ being away from zero, say $|J| > 0.1$. In Fig. \ref{f_(0,2)_(1,0)} it is shown that $f(k)$ exhibits large variations in a vicinity of $k_c=\pi/2$ and in this region it is sensitive to the value of $c$; for $c=1$, $f(k)$ exhibits a discontinuity at $\pi/2$. Far away from $\pi/2$ the function $f(k)$ is insensitive to $c$.  The plot of $-\ln f(k)$ in doubly logarithmic scale (right panel of Fig. \ref{h_(0,2)_(1,0)}) reveals the $-\ln f(k) \sim \kappa^{-2}$ behavior ($\kappa=\pi/2 -k$), sufficiently far away from $\pi/2$.
Since for given $\delta$ and sufficiently small system size $L$, the grid of wave numbers does not sample well the vicinity of $\pi/2$, where $f(k)$ varies strongly and is sensitive to $c$ (the closest to $\pi/2$ points of the grid, $\pi/2 \pm \pi/L$ are beyond this vicinity), $-\ln{\mathscr{F}}_{(1,0)}((0,J), \delta))$ does not depend on $c$ and scales with the system size as $L^2$. This is  the small-system behavior. In Fig. \ref{Fid_(0,2)_(1,0)}, where $L$ is fixed, this behavior is well visible for sufficiently small $\delta$.
To get an approximate formula for $-\ln{\mathscr{F}}_{(1,0)}((0,J), \delta))$ in the small-system regime, in formula (\ref{f_(0,J)_(1,0)}) we approximate $\cos k$ by $\kappa$ and expand in $\delta$ at zero, keeping only the lowest order term:
\begin{equation}
-\ln f(\pi/2 - \kappa) \approx \left( \frac{J}{1+J^2} \frac{\delta}{\kappa} \right)^2,
\label{f_(0,J)_(1,0)_ss}
\end{equation}
and then we sum contributions from the two closest to $\pi/2$ wave numbers $\kappa=\pm \pi/L$:
\begin{equation}
-\ln{\mathscr{F}}_{(1,0)} \left( (0,J), \delta \right) \approx \frac{2}{\pi^2} \frac{J^2}{(1+J^2)^2}(\delta L)^2.
\label{F (0,J) (1,0) lJ ss}
\end{equation}
Formula (\ref{F (0,J) (1,0) lJ ss}) approximates well $-\ln{\mathscr{F}}_{(1,0)} \left((0,J), \delta \right)$ in the whole  region of $\delta$, where fidelity is independent of $c$, i.e. in the small-system regime,
see Fig. \ref{Fid_(0,2)_(1,0)} (bottom left).

On the other hand, for given $\delta$ and sufficiently large $L$ the grid of wave numbers samples densely the vicinity of $\pi/2$, where $f(k)$ varies strongly and is sensitive to $c$.
Consequently, $-\ln{\mathscr{F}}_{(1,0)}((0,J), \delta))$ is sensitive to $c$ and scales with the system size as $L$. This is the macroscopic-system behavior.
In Fig. \ref{Fid_(0,2)_(1,0)}, where $L$ is fixed, this behavior is well visible for sufficiently large $\delta$.

The small-macroscopic system crossover is given by those $\delta$ and $L$ that satisfy the crossover condition (\ref{crossover}), which on substituting (\ref{xi}) assumes the form
\begin{equation}
|\delta| L \sim  \frac{1+J^2}{(c+1)|J|},
\label{m0_ss_ms_cross}
\end{equation}
and for sufficiently large $|J|$ simplifies to
\begin{equation}
|\delta| L \sim  \frac{|J|}{c+1}.
\label{m0_ss_ms_cross_lJ}
\end{equation}
Numerical calculations show that crossover condition (\ref{m0_ss_ms_cross_lJ}) is fairly well satisfied.

In the macroscopic-system regime, we have been able to find analytical approximations to $-\ln{\mathscr{F}}_{(1,0)}((0,J), \delta))$ for $c>1$ and for $c=0$. Specifically, for $c>1$
\begin{equation}
-\ln{\mathscr{F}}_{(1,0)}((0,J), \delta)) \approx \frac{|\delta|  L}{4c|J|} ,
\label{F (0,J) (1,0) lJ ms c>1}
\end{equation}
and for $c=0$,
\begin{equation}
-\ln{\mathscr{F}}_{(1,0)}((0,J), \delta)) \approx
\frac{-|\delta| L}{2\pi(1+J^2)}\left[ \left( 2+J^2 \right) \ln\left(1+\frac{4}{J^2} \right) +
2|J| \left( \pi- \arctan\frac{2}{|J|}\right) \right].
\label{F (0,J) (1,0) lJ ms c=0}
\end{equation}
For $0 < c \leq 1$ we have not been able to find approximate formulae analogous to the above ones.
Formulae (\ref{F (0,J) (1,0) lJ ms c>1}) and (\ref{F (0,J) (1,0) lJ ms c=0}) compare well to numerical data shown in Fig. \ref{Fid_(0,2)_(1,0)}.
Finally we note that whether it is a small- or macroscopic-system regime, fidelity depends only on the product
$\delta L$, see Fig. \ref{Fid_(0,2)_(1,0)}, bottom right panel.

 Thus, near ${\bs \lambda}_c {=} (0,J)$, with $J$ away from $0$, in transitions across the $\mu {=} 0$ critical line scaling laws (\ref{susc scaling 3}) and (\ref{fid scal}), together with crossover condition (\ref{crossover}), (\ref{m0_ss_ms_cross_lJ}) set by $\xi$
(\ref{xi}) are obeyed.

\begin{figure}\centering
\begin{subfigure}[b]{0.48\textwidth}\centering
\includegraphics[width=8cm,clip=on]{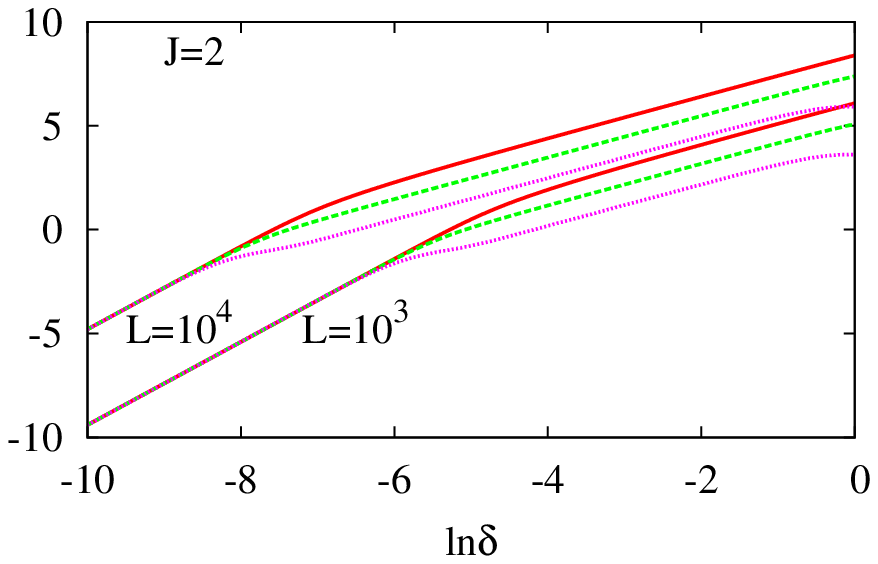}
\end{subfigure}
~
\begin{subfigure}[b]{0.48\textwidth}\centering
\includegraphics[width=8cm,clip=on]{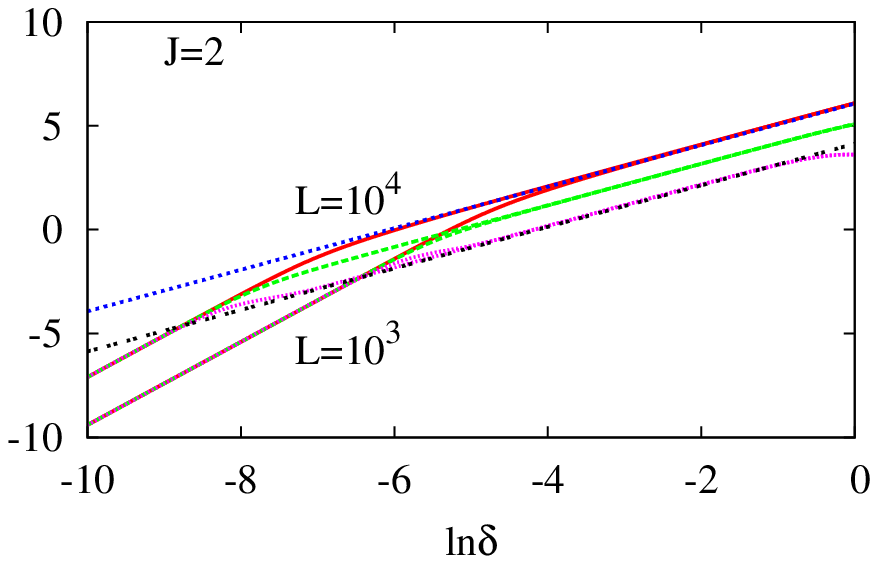}
\end{subfigure}
~
\begin{subfigure}[b]{0.48\textwidth}\centering
\includegraphics[width=8cm,clip=on]{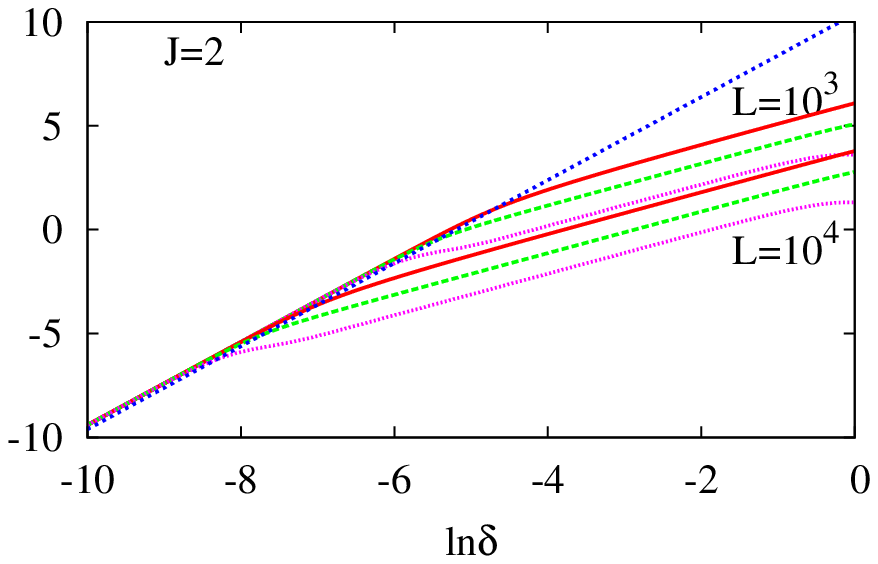}
\end{subfigure}
~
\begin{subfigure}[b]{0.48\textwidth}\centering
\includegraphics[width=8cm,clip=on]{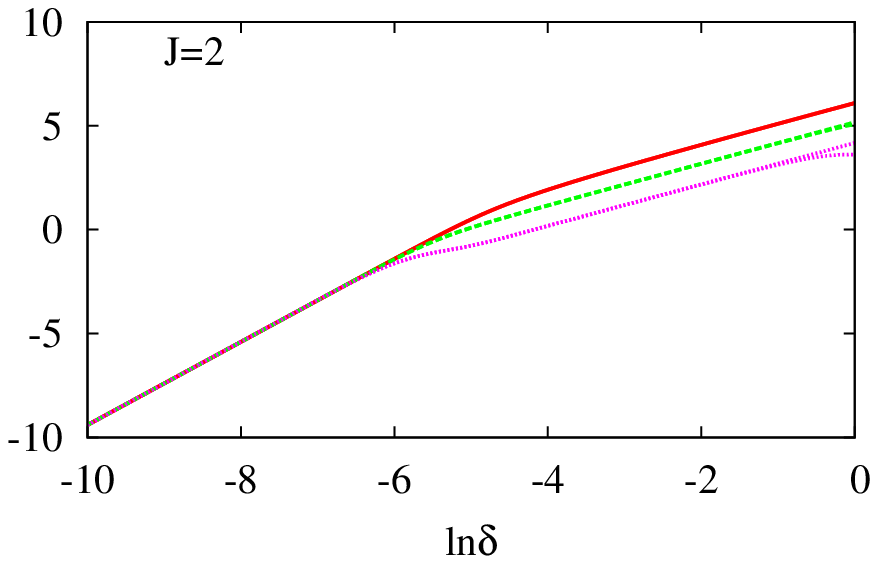}
\end{subfigure}
\caption{\label{Fid_(0,2)_(1,0)} (Color online). The plots of
$-\ln{\mathscr{F}}_{(1,0)}((0,2), \delta )$ versus $\delta$ in doubly
logarithmic scale, for two system sizes, $L=10^3$ and $L=10^4$, and three
values of $c$; for each system size, from bottom to top: $c=2$ -- magenta
dotted line, $c=1$ -- green dashed line, $c=0$ -- red line.
In top right panel the data for $L=10^4$ have been divided by 10. As a result, for
sufficiently large $\delta$ (i.e. in the macroscopic-system regime) the plot
for $L=10^4$ collapsed onto that for $L=10^3$. The blue dotted and black dotted
straight lines represent analytical approximations (\ref{F (0,J) (1,0) lJ ms
c=0}) and (\ref{F (0,J) (1,0) lJ ms c>1}), respectively, which agree well with
numerical data in the macroscopic-system regime.
In bottom left panel the data for $L=10^4$ have been divided by $10^2$ and for sufficiently small $\delta$
(i.e. in the small-system regime) the plot for $L=10^4$ collapsed onto that for
$L=10^3$. Here, the blue dotted straight line represents analytical
approximation (\ref{F (0,J) (1,0) lJ ss}), which again agrees well with
numerical data in the small-system regime.
In bottom right panel, in the plot for $L=10^4$ variable $\delta$ has been divided by $10$ which resulted
in collapse onto the plot for $L=10^3$.
Scaling is consistent with $\nu=1$, i.e. 
$-\ln{\mathscr{F}}_{(1,0)}((0,J), \delta))\sim\delta^2L^2$ in small-system regime and 
$-\ln{\mathscr{F}}_{(1,0)}((0,J), \delta))\sim |\delta| L$ in macroscopic-system regime.}
\end{figure}

%\pagebreak[4]

\subsection{\label{sJ} Across the $\mu {=} 0$ critical line: ${\bs \lambda}_c {=} (0,J)$, $J$ close to $0$, $\bs e {=} (1,0)$}

In this section, we consider $J$ being close to zero, say $|J|< 0.1$, so the critical point ${\bs \lambda}_c = (0,J)$ is in a close vicinity of  the multicritical point $(0,0)$, and we consider only $c>0$. As in the case considered in previous section \ref{lJ}, $k_c=\pi/2$, $f(k)$ is given by formula  (\ref{f_(0,J)_(1,0)}), but the character of $f(k)$ changes significantly: the plots of $f(k)$ acquire well-like shapes, see Fig. \ref{f_(0,0.01)_(1,0)}, with the positions of the "wells" depending weakly on $J$. To reveal further properties of $f(k)$ we make the corresponding plots of $-\ln f(k)$  (right panel of Fig. \ref{h_(0,0.01)_(1,0)}). Away from $\pi/2$ one can see, as in section \ref{lJ}, $-\ln f(k) \sim \kappa^{-2}$ behavior, which implies $-\ln{\mathscr{F}}_{(1,0)}((0,J), \delta ) \sim L^2$ scaling, for sufficiently small $\delta$.
In the small-system regime, fidelity is described approximately by formula (\ref{F (0,J) (1,0) lJ ss}) obtained in  section \ref{lJ}; the quality of this approximation is shown in Fig. \ref{F_(0,0.01)_(1,0)}, bottom left panel.
We note that in distinction to the situation encountered in section \ref{lJ}, the points of maximum curvature of the plots of $-\ln f(k)$ (Fig. \ref{h_(0,0.01)_(1,0)}) depend weakly on $J$. Another view of the plot of $-\ln f(k)$ for $c>1$, with some cut-off of extremely small values, is shown in Fig. \ref{h_(0,0.01)_(1,0)_c5}, for a sequence of parameter $\delta$ values. For given $L$ and sufficiently small $\delta$ the width of the "dome" is much smaller than $2\pi/L$ that separates consecutive wave numbers and its base does not incorporate any wave numbers (meaning that the contributions $-\ln f(k)$ to $-\ln{\mathscr{F}}_{(1,0)}((0,J), \delta )$ are negligible). With increasing $\delta$ the "dome" gets wider and "moves" towards the closest to $\pi/2$ wave number
$k_1=\pi/2 + \pi/L$. When the dome "passes" over $k_1$, the contribution $-\ln f(\pi/2 + \pi/L)$ increases, reaches a maximum and decreases to a very small value. As a result, a first oscillation with pronounced maximum appears in the plot of $-\ln{\mathscr{F}}_{(1,0)}((0,J), \delta )$ for $c=5$, see Fig. \ref{F_(0,0.01)_(1,0)}. Then, this scenario is repeated but for the next wave number $k_2=\pi/2 + 3\pi/L$; consequently the second maximum in those plots develops. With increasing $\delta$ the "dome" base gets wider than the separation of wave numbers, $2\pi/L$, so more than one wave number can fall under the "dome". Therefore, the oscillations become less pronounced, their amplitude decreases, eventually  a simple  $-\ln{\mathscr{F}}_{(1,0)}((0,J), \delta ) \sim \delta$ scaling results for sufficiently large $\delta$.

We can easily determine the locations in $\delta$ and the values of the described maxima of $-\ln{\mathscr{F}}_{(1,0)}((0,J), \delta )$.
For sufficiently small $|\delta|$, maxima of $-\ln f(k)$ (in continuous variable $k$) are attained at the points
\begin{equation}
k_{max}= \pi/2 \pm \sqrt{\frac{c^2 - 1}{1 + J^2}} \delta,
\label{lnf_maxima}
\end{equation}
and their value is
\begin{equation}
-\ln f(k_{max}) = -\ln \frac{1}{2} \left( 1 +
\frac{\sqrt{(c^2-1)(1+J^2)-c}}{c\sqrt{1+J^2}-\sqrt{c^2-1}} \right).
\label{lnf_maximum_value}
\end{equation}
Therefore, for finite $L$ the maxima of $-\ln{\mathscr{F}}_{(1,0)}((0,J), \delta )$ are attained at the points $\delta_n$
\begin{equation}
\delta_n = \pm \sqrt{\frac{1 + J^2}{c^2 - 1}} \frac{(2n+1)\pi}{L}, \quad n=0,1,\ldots,
\label{Fid_maxima}
\end{equation}
see Fig. \ref{F_(0,0.01)_(1,0)}, bottom right panel.
In the macroscopic-system regime, for $J$ close to $0$, we obtained analytical approximations to $-\ln{\mathscr{F}}_{(1,0)}((0,J), \delta )$ for different $c$. For $c=0$  formula (\ref{F (0,J) (1,0) lJ ms c=0}) of the previous section holds. Then, for $c=1$
\begin{equation}
-\ln{\mathscr{F}}_{(1,0)}((0,J), \delta ) \approx \frac{L|\delta|}{\pi} \left[ \ln \frac{4}{J^2}  +\frac{5}{6}J^2 - \frac{3}{2} \right],
\label{F (0,J) (1,0) sJ ms c=1}
\end{equation}
and for $c>1$
\begin{equation}
-\ln{\mathscr{F}}_{(1,0)}((0,J), \delta ) \approx
\frac{L|\delta|}{\pi} \left[ \ln \left( 1+\frac{4}{J^2c^2} \right) + |J|c
\arctan \frac{2}{|J|c}-2 \right].
\label{F (0,J) (1,0) sJ ms c>1}
\end{equation}
The plots obtained from formulae (\ref{F (0,J) (1,0) lJ ms c=0}), (\ref{F (0,J) (1,0) sJ ms c=1}) and
(\ref{F (0,J) (1,0) sJ ms c>1}) compare well with corresponding plots obtained numerically,
see Fig. \ref{F_(0,0.01)_(1,0)}, top right panel.

The plots in Fig. \ref{F_(0,0.01)_(1,0)} make clear that the location of the small- macroscopic-system crossover region can be identified with the location of the first abrupt increase in  the plot of
$-\ln{\mathscr{F}}_{(1,0)}((0,J), \delta )$ versus $\delta$. We have verified that this abrupt increase is determined by the position of the  point of maximum curvature in the plots of $-\ln f(k)$. For $c>1$, it is the location of the first maximum in  the plot of $-\ln{\mathscr{F}}_{(1,0)}((0,J)$, that can be identified with the crossover region.
Therefore, the crossover condition for the transition from $(\delta L)^2$ scaling to $|\delta| L$ scaling of $-\ln{\mathscr{F}}_{(1,0)}((0,J), \delta )$ reads
\begin{equation}
|\delta| L \sim \sqrt{\frac{1 + J^2}{c^2 - 1}}\pi.
\label{m0_ss_ms_cross_sJ}
\end{equation}
For sufficiently small $|J|$, when the considered critical point is in close proximity to the multicritical point $(0,0)$, the crossover condition (\ref{m0_ss_ms_cross_sJ}) is almost independent of $J$. Formula (\ref{m0_ss_ms_cross_sJ}) is in contrast with the expected condition, given by formula (\ref{m0_ss_ms_cross}), which for sufficiently small $|J|$ reads:  $|\delta| L \sim 1/(c+1)|J|$.
This result can be interpreted as the appearance of a new,  $J$-independent for sufficiently small $|J|$, characteristic length in the system, which we denote $\xi'$. The appearance of  $\xi'$  can naturally be attributed to the influence of the multicritical point.

 Thus, near ${\bs \lambda}_c {=} (0,J)$, with $J$ close to $0$, in transitions across the $\mu {=} 0$ critical line scaling laws (\ref{susc scaling 3}) and (\ref{fid scal}) are obeyed but crossover condition (\ref{crossover}), (\ref{m0_ss_ms_cross_lJ}) set by $\xi$
(\ref{xi}) is not.
\begin{figure*}\centering
\begin{subfigure}[b]{0.48\textwidth}\centering
\includegraphics[width=8cm,clip=on]{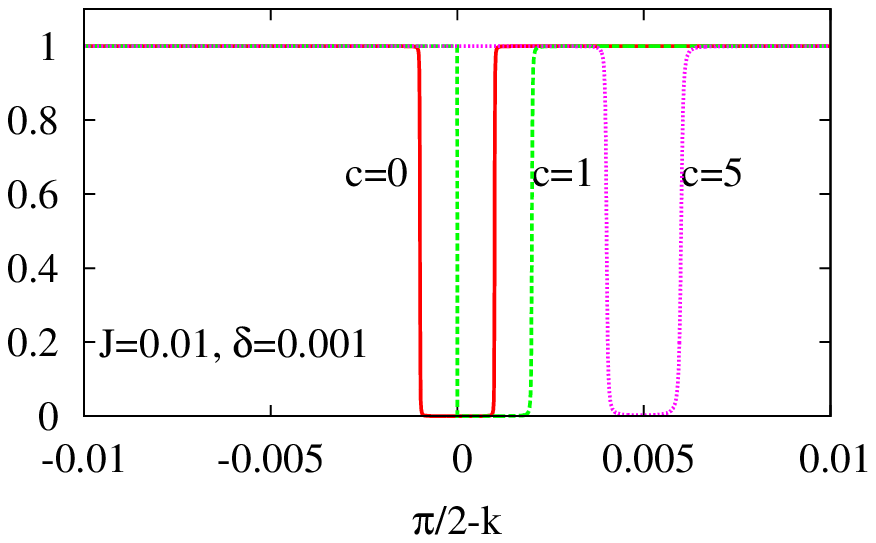}
\end{subfigure}
~
\begin{subfigure}[b]{0.48\textwidth}\centering
\includegraphics[width=8cm,clip=on]{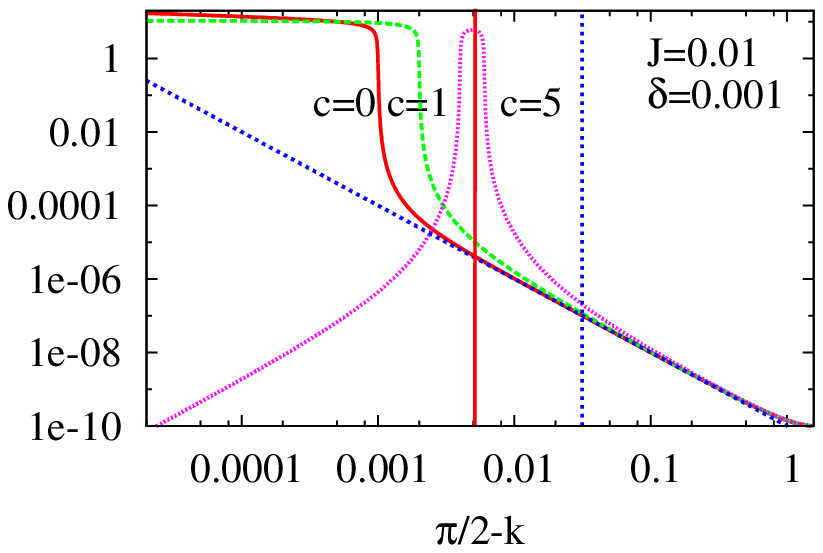}
\end{subfigure}
\caption{\label{f_(0,0.01)_(1,0)}\label{h_(0,0.01)_(1,0)}(Color online). The
plots of $f$ (left panel) and $-\ln f$ (right panel) versus $\pi/2 - k$ for three values of
$c$: $c=0$ -- red solid line, $c=1$ -- green dashed line, $c=5$ -- magenta
dotted line. Right panel is in doubly logarithmic scale; the
blue dotted straight line of slope $-2$  indicates the $-\ln f(k) \sim
\kappa^{-2}$ behavior, red and blue vertical lines indicate the location of the
wave number $\pi/2 - \pi/L$ that is the closest to $\pi/2$ for $L=600$ and
$L=100$, respectively.}
\end{figure*}

\begin{figure}
\begin{center}
\includegraphics[width=8cm,clip=on]{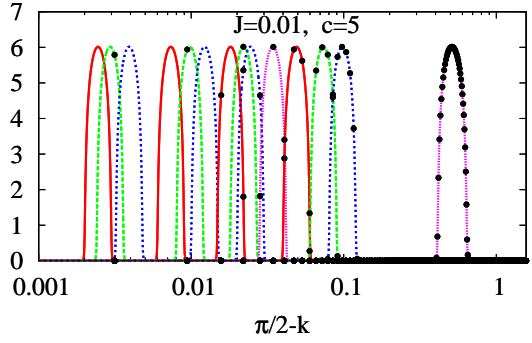}
\caption{\label{h_(0,0.01)_(1,0)_c5} (Color online). For $J=10^{-2}$, $c=5$, $L=10^3$, the plot of $-\ln f$  versus $\pi/2 - k$ (abscissa in logarithmic scale) for  a sequence of $\delta$; from left to right, $\delta 10^4=$ $5, 6, 8, 15, 20, 25, 37, 45, 50, 70, 100, 150, 200, 1000$. For each $\delta$ the plot is of steep-dome-like shape whose base width is approximately $2\delta$. The black points on abscissa mark the position of wave numbers, which are separated by $2\pi/L$. The black points on a dome indicate how many wave numbers are included in its base and what is their contribution to fidelity. }
\end{center}
\end{figure}

\begin{figure*}\centering
\begin{subfigure}[b]{0.48\textwidth}\centering
\includegraphics[width=8cm,clip=on]{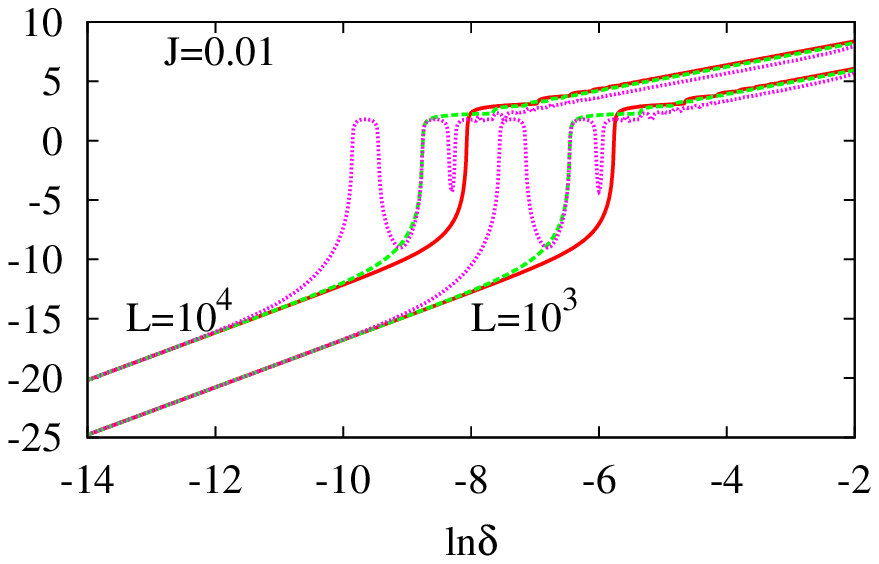}
\end{subfigure}
~
\begin{subfigure}[b]{0.48\textwidth}\centering
\includegraphics[width=8cm,clip=on]{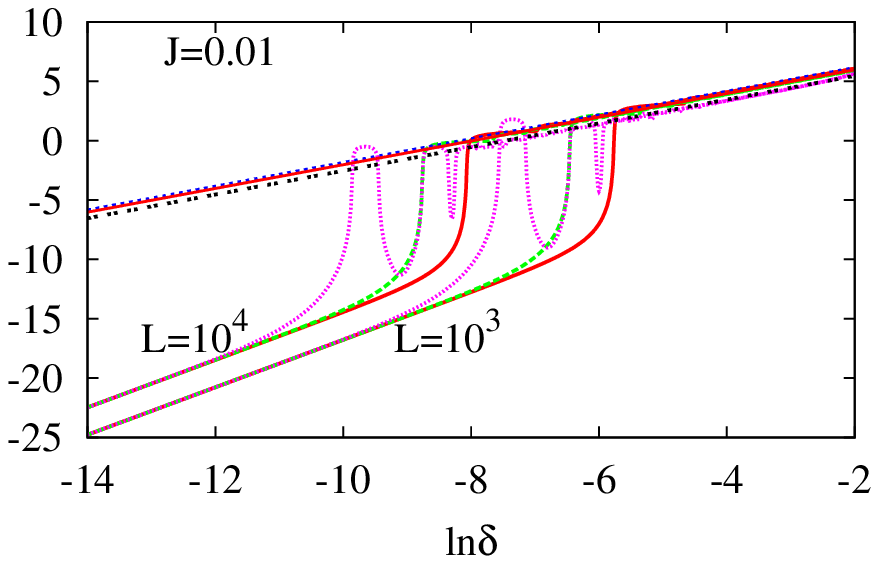}
\end{subfigure}
~
\begin{subfigure}[b]{0.48\textwidth}\centering
\includegraphics[width=8cm,clip=on]{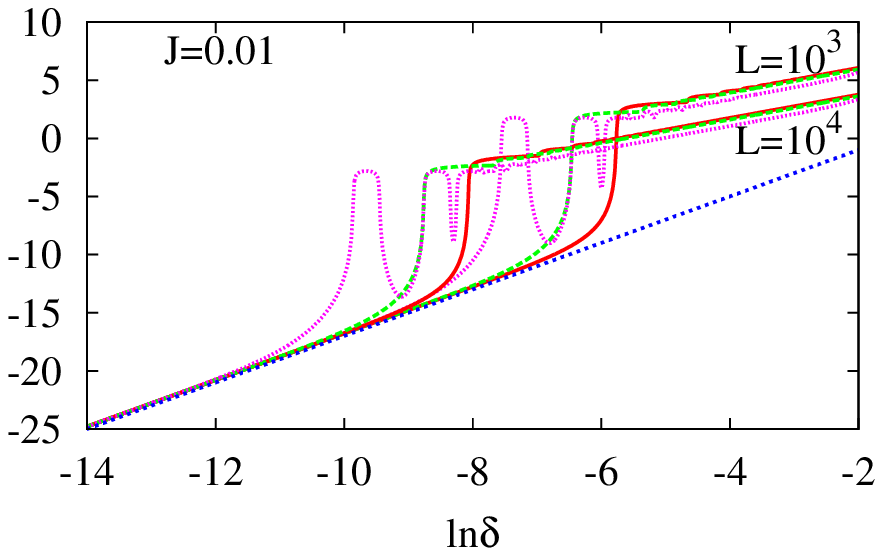}
\end{subfigure}
~
\begin{subfigure}[b]{0.48\textwidth}\centering
\includegraphics[width=8cm,clip=on]{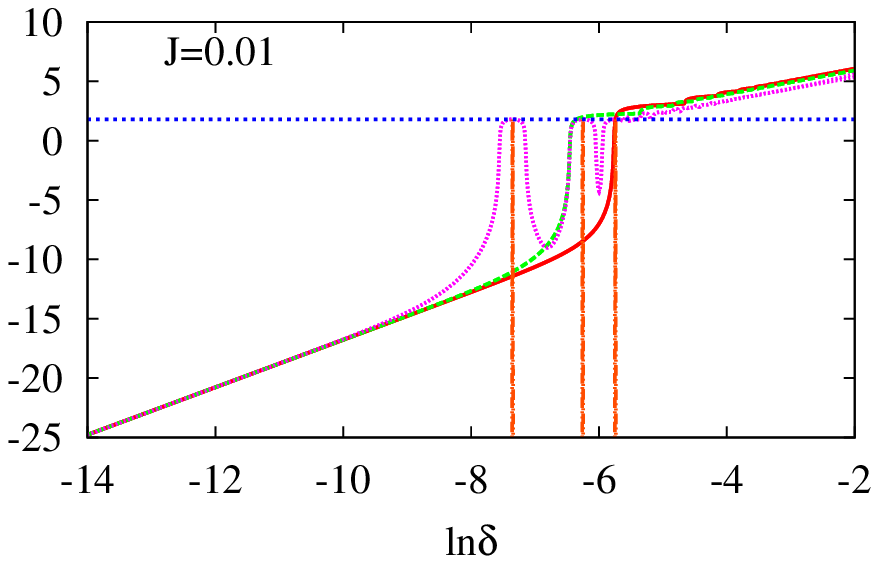}
\end{subfigure}
\caption{\label{F_(0,0.01)_(1,0)} (Color online). Plots of
$-\ln{\mathscr{F}}_{(1,0)}((0,10^{-2}), \delta)$  versus $\delta$ in doubly
logarithmic scale, for two system sizes, $L=10^3$ and $L=10^4$, and three
values of $c$; for each system size, from left to right: $c=5$ -- magenta
dotted line, $c=1$ -- green dashed line, $c=0$ -- red line. In top right panel
the data for $L=10^4$ have been divided by 10. As a result, for sufficiently
large $\delta$ (i.e. in the macroscopic-system regime) the plots for $L=10^4$
and for $L=10^3$ merged into one curve.  The straight lines in the upper part of
the figure represent analytical approximations: formula (\ref{F (0,J) (1,0) lJ
ms c=0}) for $c=0$ -- blue dotted line, formula (\ref{F (0,J) (1,0) sJ ms c=1})
for $c=1$ -- red continuous line, and formula (\ref{F (0,J) (1,0) sJ ms c>1})
for $c=5$ -- black dotted line. All the three analytical approximations agree
well with numerical data for sufficiently large $\delta$, i.e. in the
macroscopic-system regime. In bottom left panel the data for $L=10^4$ have been
divided by $10^2$. As a result, for sufficiently small $\delta$ (i.e. in the
small-system regime) the plots for $L=10^4$ and $L=10^3$ merged into one curve.
The blue dotted straight line represents analytical approximation
(\ref{F (0,J) (1,0) lJ ss}), which agrees well with numerical data in the
small-system regime. In bottom right panel, variable $\delta$ has been divided
by $10$. As a result, the plots for $L=10^4$ and $L=10^3$ merged into one curve.
The blue dotted line indicates that the maxima of the oscillations in
the plot for $c=5$  are of the same height, given by (\ref{lnf_maximum_value}),
and are attained at points $\delta_n$, given by (\ref{Fid_maxima}). The orange
vertical lines mark the positions of the maxima on the $\delta$-axis. These
positions are: $\delta 10^4$=6.3, 19.3, 32.1 (not shown), and 37.
Scaling is consistent with $\nu=1$, 
i.e. $-\ln{\mathscr{F}}_{(1,0)}((0,J), \delta )\sim\delta^2L^2$ in small-system regime
and $-\ln{\mathscr{F}}_{(1,0)}((0,J), \delta )\sim|\delta| L$ in macroscopic-system regime.}
\end{figure*}

\subsection{\label{Jaxis} Along the $J$-axis: ${\bs \lambda}_c {=} (0,J)$, arbitrary $J$, $\bs e {=} (0,1)$}

Along the $J$-axis, that is a critical point and its neighborhood are contained in $J$-axis, we encounter a particularly simple situation. Since $\mu=\tilde{\mu}=0$, $f(k)$ simplifies to a constant:
\begin{eqnarray}
f(k)= \frac{1}{2} \left \{ 1 +
\frac{1 + \left[ J + (c-1) \delta \right] \left[ J + (c+1) \delta \right]}
{\sqrt{ \left( 1+ \left[ J + (c-1) \delta \right]^2  \right) \left( 1+ \left[ J + (c+1) \delta \right]^2 \right) }}
\right \}.
\label{(0,J)_f_(0,1)}
\end{eqnarray}
Expanding $-\ln f(k)$ in $\delta$ at zero we obtain
\begin{equation}
-\ln{\mathscr{F}}_{(0,1)}((0,J), \delta) = \frac{1}{2} \left(\frac{\delta}{1+J^2}\right)^2 L + {\cal{O}}(\delta^3) L.
\label{F_(0,J) (0,1)}
\end{equation}
Taking into account only the lowest order in $\delta$ term, we obtain an accurate approximation to $-\ln{\mathscr{F}}_{(0,1)}((0,J), \delta)$.  The anomalous scaling (\ref{F_(0,J) (0,1)}) holds in a wide range of values of parameter $\delta$.
As shown in Fig. \ref{Fid_(0,2)_(0,1)}, this range covers the whole critical region, including the small and macroscopic system regimes, there is no small-system - macroscopic-system crossover.
The plots of $-\ln{\mathscr{F}}_{(0,1)}((0,J), \delta)$ versus $\delta$,
obtained for different values of $J$ or $L$, after suitable rescaling according to (\ref{F_(0,J) (0,1)}), merge into  one curve.

 Thus, in transitions along $J$-axis, near ${\bs \lambda}_c {=} (0,J)$, there is no small-system - macroscopic system crossover and scaling laws  (\ref{susc scaling 3}) and (\ref{fid scal}) are not obeyed.

\begin{figure}[ht]
\begin{center}
\includegraphics[width=8cm,clip=on]{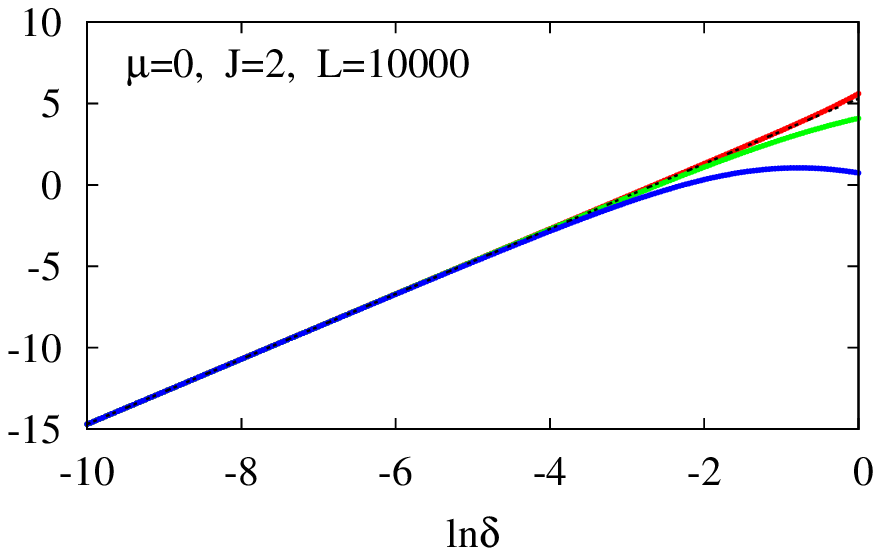}
\caption{\label{Fid_(0,2)_(0,1)} (Color online). $-\ln{\mathscr{F}}_{(0,1)}((0,J), \delta)$ versus $\delta$ in doubly logarithmic scale, for different $c$; from bottom to top: $c=5$ -- blue line, $c=1$ -- green line, $c=0$ -- red line. The black dotted straight line of slope $2$, coinciding everywhere with the red line and for $\ln \delta < -4$ also with green and blue lines, represents the first term of formula (\ref{F_(0,J) (0,1)}).
Results consistent with infinite $\xi$.}
\end{center}
\end{figure}

%\clearpage

\subsection{\label{mcp} The multicritical point ${\bs \lambda}_c {=} (0,0)$, various directions $\bs e$}

At the multicritical point the gap $E_k$ closes at $k_c=\pi/2$.
In a line neighborhood in direction ${\bs e}= (\cos\phi,\sin\phi)$, where $0 < \phi < \pi/2$, of  the multicritical point the function $f(k)$ assumes the form
\begin{eqnarray}
f(k)=\frac{1}{2}\left \{ 1+\frac{(\cos k- \mu)(\cos k -\tilde{\mu})+J\tilde{J} \cos^2k}
{\sqrt{ \left[ (\cos k -\mu)^2+J^2\cos^2 k \right] \left[(\cos k -\tilde{\mu})^2+\tilde{J}^2\cos^2 k\right]}}
\right \},
\label{(0,0)_f_phi}
\end{eqnarray}
where we set
\begin{eqnarray}
\mu=(c-1)\delta\cos\phi,\  \tilde{\mu}=(c+1)\delta\cos\phi, \qquad
J=(c-1)\delta\sin\phi,\  \tilde{J}=(c+1)\delta\sin\phi,
\end{eqnarray}
with $c>0$.

\begin{figure*}\centering
\begin{subfigure}[b]{0.48\textwidth}\centering
\includegraphics[width=8cm,clip=on]{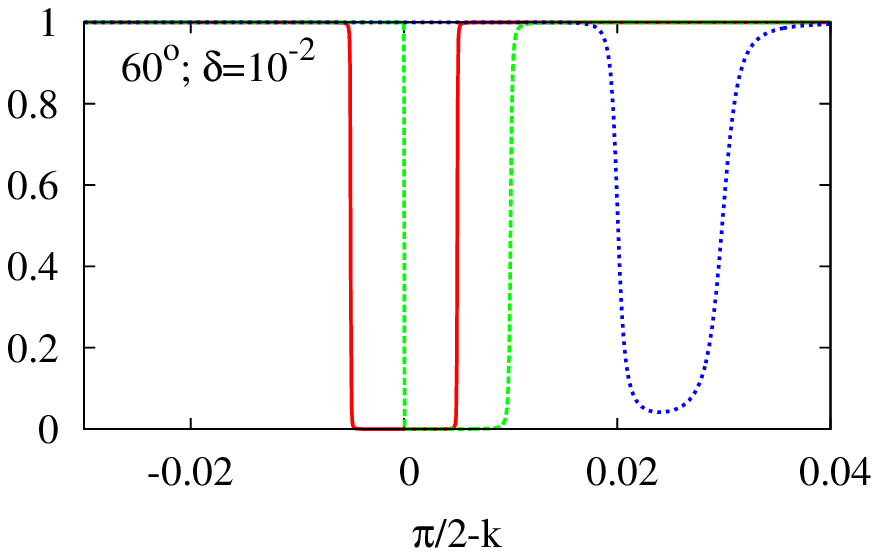}
\end{subfigure}
~
\begin{subfigure}[b]{0.48\textwidth}\centering
\includegraphics[width=8cm,clip=on]{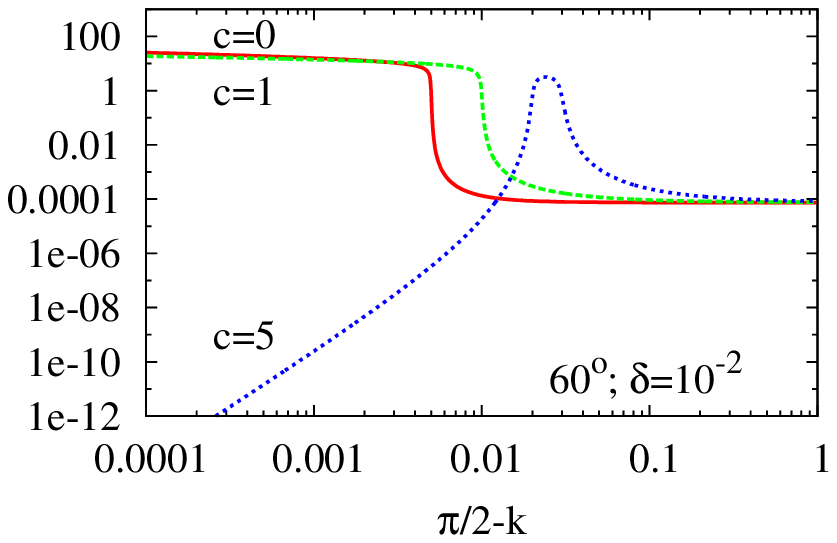}
\end{subfigure}
\caption{\label{f_(0,0)_60}\label{h_(0,0)_60} (Color online) The plots of
$f$ (left panel) and $-\ln f$ (right panel) versus $\kappa=\pi/2 - k$ for $\phi=60^o$ and three values
of $c$: $c=0$ -- red solid line, $c=1$ -- green dotted line and, $c=5$ -- blue
dotted line. Right panel is in doubly logarithmic scale.}
\end{figure*}

\begin{figure}
\begin{center}
\includegraphics[width=8cm,clip=on]{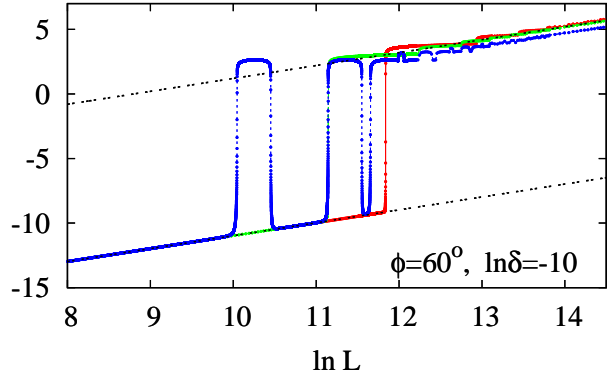}
\caption{\label{Fid_phi60_od_L} (Color online)
$-\ln {\mathscr{F}}_{(\cos 60^o,\sin 60^o)}((0,0), e^{-10})$ versus $L$, in doubly logarithmic scale, for three values of $c$; from left to right: $c=5$ -- blue line, $c=1$ -- green line, and $c=0$ -- red line.
Two black dotted straight lines of slope $1$ indicate that in the small system and macroscopic system regimes
$-\ln {\mathscr{F}}_{(\cos 60^o,\sin 60^o)}((0,0), e^{-10}) $ scales with the system size as $L$. }
\end{center}
\end{figure}

\begin{figure*}\centering
\begin{subfigure}[b]{0.48\textwidth}\centering
\includegraphics[width=8cm,clip=on]{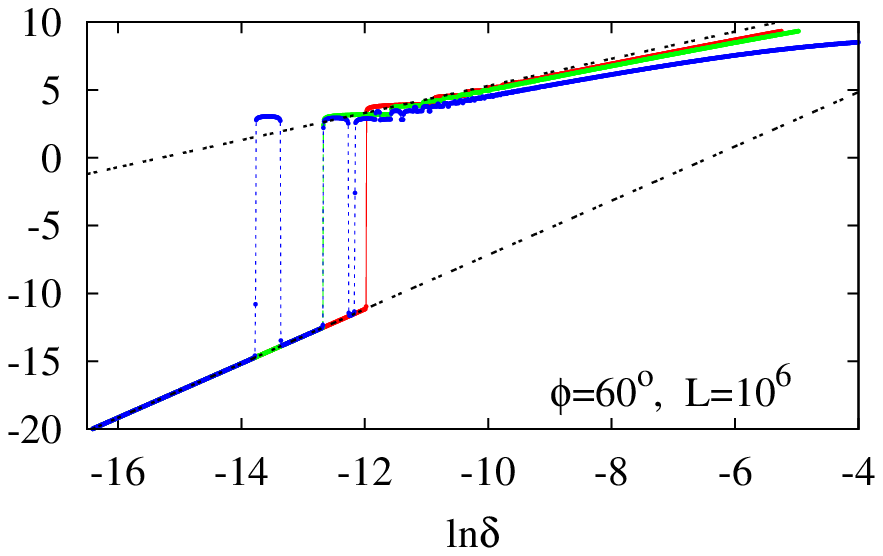}
\end{subfigure}
~
\begin{subfigure}[b]{0.48\textwidth}\centering
\includegraphics[width=8cm,clip=on]{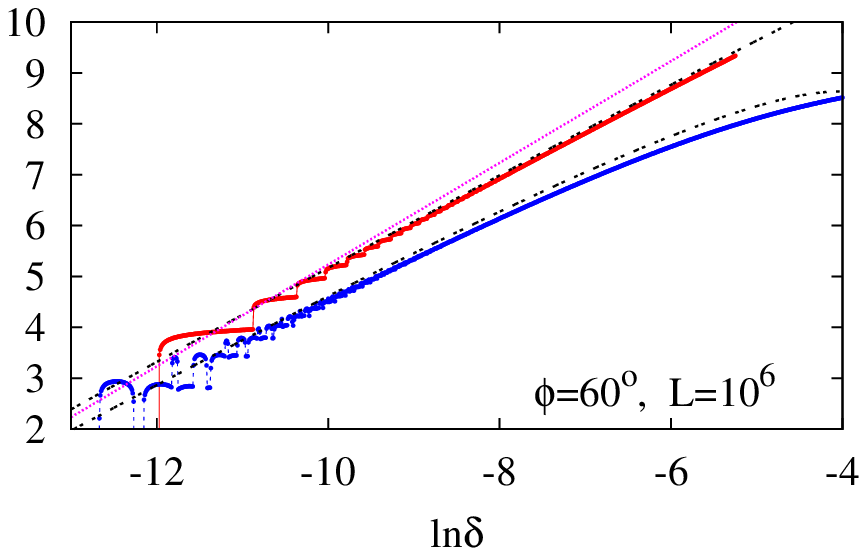}
\end{subfigure}
\caption{\label{Fid_phi60_od_d}\label{Fid_phi60_od_d2} (Color online) Plots of $-\ln
{\mathscr{F}}_{(\cos 60^o,\sin 60^o)}((0,0), \delta)$ versus $\delta$ in
doubly logarithmic scale, for system size $L=10^6$ and three values of $c$;
from left to right: $c=5$ -- blue line, $c=1$ -- green line, and $c=0$ -- red
line. In left panel, the lower black dotted straight line of slope 2 is the
plot of (\ref{(0,0)_ss_Fid}), and well agrees with numerical data in the
small-system regime. The upper black dotted straight line of slope 1 indicates
that in the macroscopic-system regime $-\ln {\mathscr{F}}_{(\cos 60^o,\sin
60^o)}((0,0), \delta) $ does not scale as a power of $\delta$. Right panel
shows the magnified view of the macroscopic-system regime. Here, the lower black
dotted line is the analytical approximation (\ref{(0,0_Fid_c_gr_1}) to the blue
line and the upper black dotted line is the analytical approximation
(\ref{(0,0)_Fid_c_less_1}) to the red line. The uppermost magenta straight line
of slope 1 is a reference line.}
\end{figure*}

\begin{figure}
\begin{center}
\includegraphics[width=8cm,clip=on]{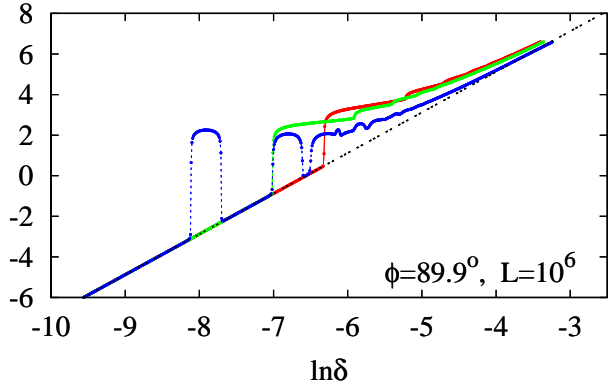}
\caption{\label{Fid_phi899_od_d} (Color online) The system size is $L=10^6$.
$-\ln {\mathscr{F}}_{(\cos 89.9^o,\sin 89.9^o)}((0,0), \delta)$ versus $\delta$, in doubly logarithmic scale,
for three values of $c$; from left to right: $c=5$ -- blue line, $c=1$ -- green line, and $c=0$ -- red line.
The lower black dotted straight line of slope 2 is the plot of (\ref{(0,0)_ss_Fid}).}
\end{center}
\end{figure}

If $\delta$ is not too large, say $\delta<10^{-2}$, the plots of $f(k)$ for different $c$  are of well-like shape (see Fig. \ref{f_(0,0)_60}). The wells are narrow and located close to $k_c=\pi/2$; their widths, calculated at the level $f(k)=1/2$, are proportional to $\delta \cos \phi$, as a simple calculation reveals. Moreover, looking at the plots in Fig.~\ref{f_(0,0)_60} it is clear that for small systems, if there are no wave numbers inside the wells, $f(k)$ equals approximately $1$.
Therefore, to get an approximate analytic expression for $-\ln {\mathscr{F}}_{\bs e}({\bs{\lambda}}_c, \delta)$ in the small-system regime we expand $f(k)$ in $\delta$ at zero, keeping only the lowest order term, which is $k$-independent,
\begin{eqnarray}
-\ln {\mathscr{F}}_{(\cos \phi,\sin \phi)}((0,0), \delta) \approx \frac{\sin^2 \phi}{2} \delta^2 L.
\label{(0,0)_ss_Fid}
\end{eqnarray}
This formula compares very well with numerical data shown in Figs. \ref{Fid_phi60_od_L}, \ref{Fid_phi60_od_d}, and \ref{Fid_phi899_od_d}.
 Clearly, small-system scaling law (\ref{susc scaling 3}) is not obeyed.

For macroscopic systems, when for given $\delta$ the  system size $L$  is large enough, it is the bottom of the well that plays a dominant role. In a narrow well this bottom is quite flat, so we expand $f(k)$ in $k$ at the minimum of the well, again keeping only the lowest order term. Then, the result, which is sensitive to the value of $c$, is integrated between the two points that are solutions to the equation $f(k)=1/2$. For sufficiently small $\delta$ the approximate solutions of this equation are given by the solutions $k_1,k_2$ of equations: $\cos k_1=\tilde{\mu}$ and $\cos k_2=\mu$. For all $c$, the result of integration is of the form ${\cal{C}} |\delta| + {\cal{C}}^{\prime} |\delta| \ln |\delta|$, where ${\cal{C}}$, ${\cal{C}}^{\prime}$ depend on $c$ and $\phi$.

Specifically, for $c<1$, the minimal value of $f(k)$ is zero and is attained at $\pi/2$, and the lowest-order term of Taylor expansion at $k=\pi/2$ is
\begin{equation}
f\left(\frac{\pi}{2} - \kappa\right)\approx
\frac{1}{\delta^2}\frac{1}{(1-c^2)^2}\frac{\sin^2\phi}{\cos^4\phi}\kappa^4,
\label{(0,0)_f_c_less_1}
\end{equation}
which after integration gives
\begin{equation}
-L^{-1}\ln {\mathscr{F}}_{(\cos \phi,\sin \phi)}((0,0), \delta) \approx
{\cal{C}}_1 |\delta| + {\cal{C}}_2 |\delta| \ln |\delta|,
\label{(0,0)_Fid_c_less_1}
\end{equation}
with
\begin{eqnarray}
{\cal{C}}_1 = \frac{2 \cos\phi}{\pi} \left( 2 - c\ln \frac{1+c}{1-c} - \ln \sin\phi \right),\qquad
{\cal{C}}_2 = -\frac{2 \cos\phi}{\pi}.
\label{C1_C2_c_less_1}
\end{eqnarray}

Then, for $c=1$
\begin{equation}
f\left(\frac{\pi}{2} - \kappa\right) \approx \frac{\tan^2\phi}{4} \kappa^2,
\label{(0,0)_f_c_eq_1}
\end{equation}
therefore
\begin{equation}
-L^{-1}\ln {\mathscr{F}}_{(\cos \phi,\sin \phi)}((0,0), \delta) \approx
{\cal{C}}_3 |\delta| + {\cal{C}}_2 |\delta| \ln |\delta|,
\label{(0,0_Fid_c_eq_1}
\end{equation}
with
\begin{eqnarray}
{\cal{C}}_3 = \frac{2 \cos\phi}{\pi} \left(1 - \ln \sin\phi \right).
\label{C3_c_eq_1}
\end{eqnarray}

Finally, for $c>1$ we approximate the bottom of the well by a constant equal to the minimal value of the well, which for sufficiently small $\delta$ is  approximately equal to
\begin{equation}
\min f(k)  \approx (c^2-1)^2\delta^2\sin^2\phi.
\label{(0,0)_f_c_gr_1}
\end{equation}
Then, integration leads to
\begin{equation}
-L^{-1}\ln {\mathscr{F}}_{(\cos \phi,\sin \phi)}((0,0), \delta) \approx
{\cal{C}}_4 |\delta| + {\cal{C}}_2 |\delta| \ln |\delta|,
\label{(0,0_Fid_c_gr_1}
\end{equation}
with
\begin{eqnarray}
{\cal{C}}_4 = - \frac{2\cos\phi}{\pi} \ln \left( (c^2-1)\sin\phi \right).
\label{C4_c_gr_1}
\end{eqnarray}
In Fig. \ref{Fid_phi60_od_d2} (right panel) it is shown that, for some range of $\delta$ in the macroscopic-system regime,
formula (\ref{(0,0)_Fid_c_less_1}) well agrees with
$-\ln {\mathscr{F}}_{(\cos 60^o,\sin 60^o)}((0,0), \delta)$ numerically calculated for $c=0$ and $c=5$. Similar results have been obtained for other values of angles $\phi$ (between $0$ and $\pi/2$) and parameters $c$. However, if $\phi$ approaches $\pi/2$, then in the macroscopic-system regime $-\ln {\mathscr{F}}_{(\cos \phi,\sin \phi)}((0,0), \delta) $ tends asymptotically, with increasing $\delta$, towards the straight line describing small-system fidelity, given by (\ref{(0,0)_ss_Fid}). For $\phi$ very close to $\pi/2$, this small-system fidelity is identical to that found at the line of critical points given by $\mu=0$, see Fig. \ref{Fid_phi899_od_d}.

\begin{figure}[ht]
\begin{center}
\includegraphics[width=8cm,clip=on]{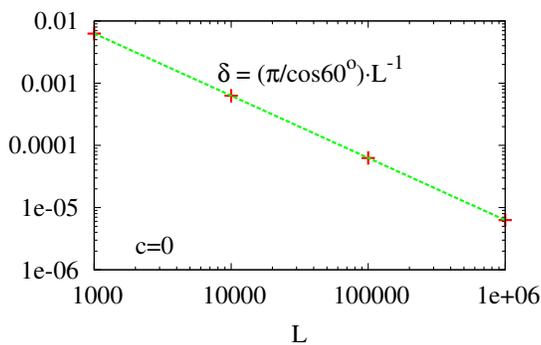}
\caption{\label{delta_v_L_60_c0}  The location in $\delta$ of the crossover region is linear in $L^{-1}$. The plot is made for $c=0$ and $\phi=60^o$. Red crosses stand for numerical data, green dashed line represents formula (\ref{Ld_cross}).}
\end{center}
\end{figure}

An inspection of figures Fig. \ref{Fid_phi60_od_L}, \ref{Fid_phi60_od_d}, \ref{Fid_phi899_od_d} shows clearly that the small- macroscopic-system crossover can readily be defined by the well-localized first "jump" in the plot of fidelity versus $\delta$ or versus $L$. Numerically we found that the position $\delta$ of the "jump" scales with system size like $L^{-1}$, for a number of values of $c$. A simple argument provides an appropriate formula -- the crossover condition. For sufficiently small $\delta$, there are no points of wave number grid inside the well in the plot of $f(k)$, so $-\ln {\mathscr{F}}_{(\cos \phi,\sin \phi)}((0,0), \delta) \sim \delta^2$. When the first point of the wave number grid enters the well, $-\ln {\mathscr{F}}_{(\cos \phi,\sin \phi)}((0,0), \delta)$ jumps to a much greater value. This occurs approximately for those $\delta$ and $L$ that satisfy the equation
$\cos (\pi/2- \kappa) = {\tilde{\mu}}$, with $\kappa=\pi/L$ and ${\tilde{\mu}} = (c+1) \delta \cos \phi$. That is, the crossover condition reads
\begin{equation}
|\delta| L \approx \pi \frac{c+1}{\cos \phi}.
\label{Ld_cross}
\end{equation}
Formula (\ref{Ld_cross}) agrees well with our numerical data, in particular with those in Fig \ref{delta_v_L_60_c0}, and is consistent with $\nu=1$.

 Summarizing scaling properties near multicritical point ${\bs \lambda}_c {=} (0,0)$, we have found that neither small-system (\ref{susc scaling 3}) nor macrosopic-system (\ref{fid scal}) scaling laws are obeyed; the dependence of  $-\ln {\mathscr{F}}_{(\cos \phi,\sin \phi)}((0,0), \delta)$ on $|\delta|$ is not power-like.

\subsection{\label{im} Across the $J {=} 0$ critical line: ${\bs \lambda}_c {=} (\mu,0)$,
 $0 {<} \mu {<} 1$, $\bs e {=} (0,1)$}

In contrast to the critical points considered in previous sections, in this case the excitation gap $E_k$ closes at $k_c$ which depends on the system parameter $\mu$; $k_c$ is the solution of the equation $\cos k_c =\mu$. This fact has profound consequences for the behavior of fidelity.
\begin{figure*}\centering
\begin{subfigure}[b]{0.48\textwidth}\centering
\includegraphics[width=8cm,clip=on]{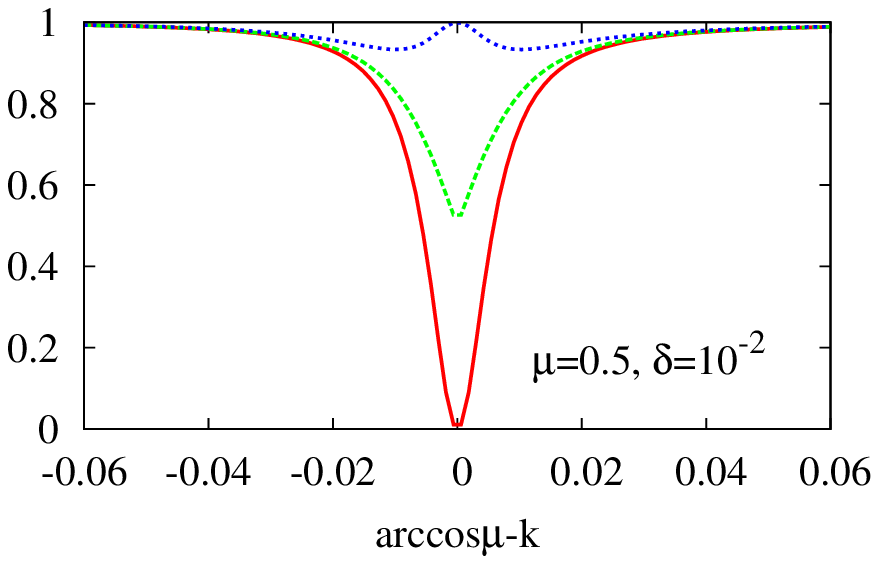}
\end{subfigure}
~
\begin{subfigure}[b]{0.48\textwidth}\centering
\includegraphics[width=8cm,clip=on]{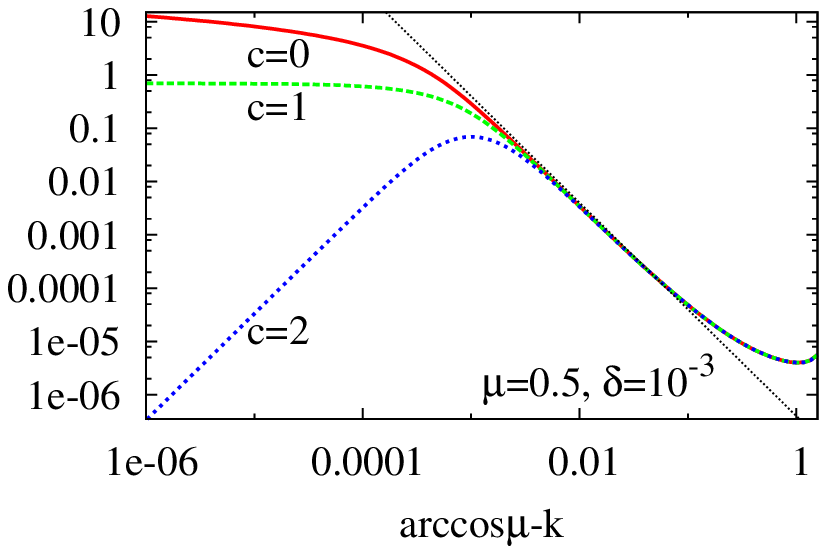}
\end{subfigure}
\caption{\label{f_(0.5,0)_(0,1)}\label{h_(0.5,0)_(0,1)} (Color online). The
plots of $f$ (left panel) and $-\ln f$ (right panel) versus $(\arccos\mu - k)$ for three values of
$c$: $c=0$ -- red solid line, $c=1$ -- green dashed line,
$c=2$ -- blue dotted line.  Right panel is in doubly logarithmic scale;
the black-dotted straight line indicates the $-\ln f(k) \sim \kappa^{-2}$ behavior.}
\end{figure*}
In a line neighborhood of a critical point $(\mu,0)$, in direction $(0,1)$, the function $f(k)$ assumes the form
\begin{eqnarray}
f(k)=\frac{1}{2}\left \{ 1+\frac{(\cos k - \mu)^2 + J\tilde{J} \cos^2k}
{\sqrt{ \left[ (\cos k - \mu)^2+J^2\cos^2 k \right] \left[(\cos k - \mu)^2 + \tilde{J}^2\cos^2 k\right]}}
\right \},
%\label{(1,0)_f}
\end{eqnarray}
where we set
\begin{eqnarray}
J=(c-1)\delta, \quad \tilde{J}=(c+1)\delta.
\end{eqnarray}
\begin{figure*}\centering
\begin{subfigure}[b]{0.48\textwidth}\centering
\includegraphics[width=8cm,clip=on]{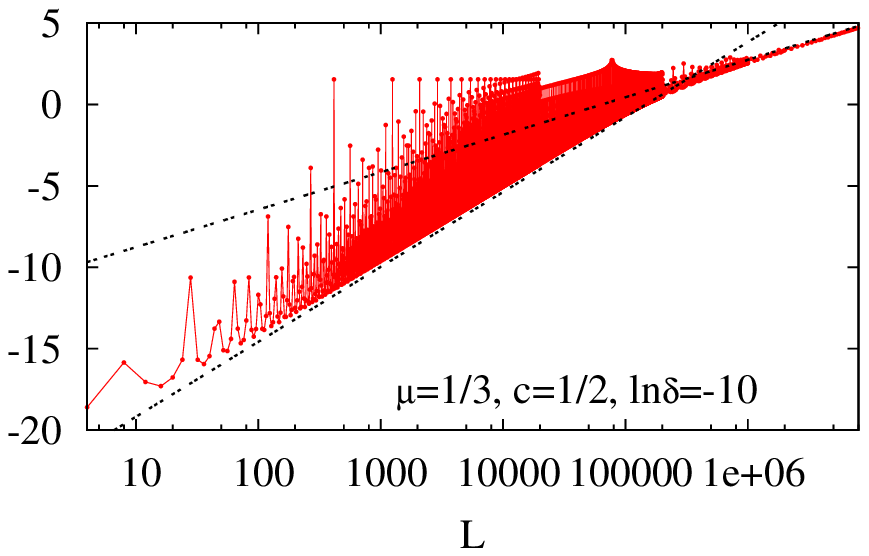}
\end{subfigure}
~
\begin{subfigure}[b]{0.48\textwidth}\centering
\includegraphics[width=8cm,clip=on]{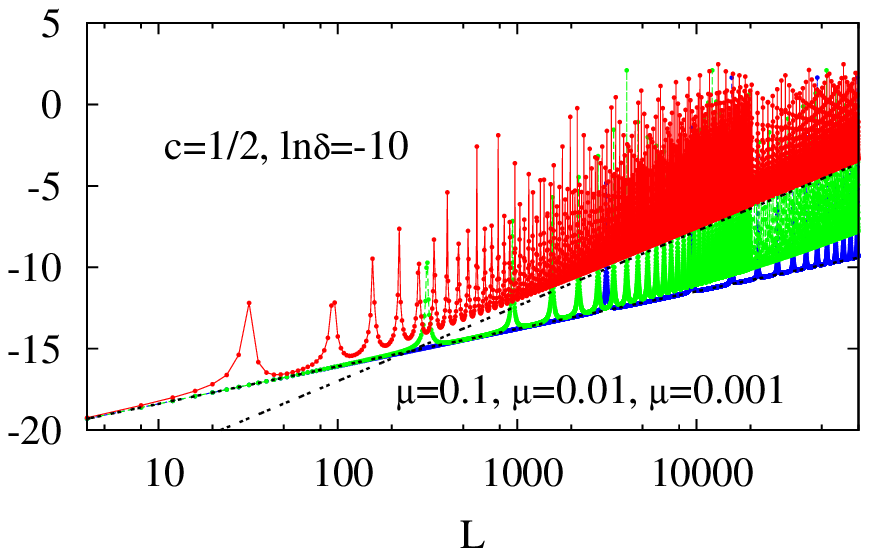}
\end{subfigure}
\caption{\label{Fid_mu0333_c05_d10_od_L}\label{Fid_mu_c05_d10_od_L} (Color
online). Left panel: plot of $-\ln {\mathscr{F}}_{(0,1)}((1/3,0), e^{-10})$ versus $L$ in
doubly logarithmic scale, for $c=1/2$. Significant oscillations whose amplitude
decreases with $L$ for sufficiently large $L$ are well visible. The black-dotted straight line of slope 2 represents formula (\ref{Fid_(m,0)_ss})with adjusted numerical coefficient. The intersection of this line with the other black-dotted straight line of slope $1$ indicates the crossover region between small-system and macroscopic-system behavior, which is manifested by the change of finite size scaling from $L^2$ to $L$.
Right panel: plot of $-\ln {\mathscr{F}}_{(0,1)}((\mu,0), e^{-10})$ versus $L$ in doubly logarithmic scale for $c=1/2$ and three values of $\mu$; from top to bottom: $\mu=10^{-1}$ -- red line, $\mu=10^{-2}$ -- green line, $\mu=10^{-3}$ -- blue line. There are two black dotted straight lines. One of them, whose slope is $1$, given by formula (\ref{F_(0,J) (0,1)}) with $J=0$, coincides with the minima of the blue line for all $L$ in the figure, and with minima of green and red lines for sufficiently small $L$. The other line, whose slope is $2$, given by formula (\ref{Fid_(m,0)_ss}) with adjusted numerical coefficient, coincides with minima of the red line for sufficiently large $L$.  The intersection of these two lines indicates, on increasing $L$, the crossover region between ultra-small-system regime with $L$-scaling and small-system regime with $L^2$-scaling.}
\end{figure*}
Plot of $f(k)$ in a neighborhood of $k_c$ is shown in  Fig. \ref{f_(0.5,0)_(0,1)}, left panel. For all $c$, the function $f(k)$ is continuous and varies rapidly only in a neighborhood of $k_c$, with the minimal value attaining zero if $|c|<1$;
therefore in the small-system regime fidelity is insensitive to $c$.
The plot of $-\ln f(k)$ in doubly logarithmic scale,  Fig. \ref{h_(0.5,0)_(0,1)}, right panel, reveals the $-\ln f(k) \sim \kappa^{-2}$ behavior ($\kappa=\arccos \mu - k$), sufficiently far away from $\arccos \mu$. However, this behavior does not imply
$-\ln {\mathscr{F}}_{(\mu,0)}((0,1), \delta) \sim L^2$ scaling in the small-system regime, as was the case in previous sections.

The point is that for a typical critical point $(\mu,0)$, when $L$ is increased the grid of wave numbers "moves" with respect to $k_c$. That is,  if the points of the grid are initially, for some $L$, distributed symmetrically about $k_c$, with increasing $L$ their distribution becomes more and more asymmetric, and there is a wave number approaching $k_c$ arbitrarily close even for finite $L$. Further increase of $L$ restores a symmetric distribution, and so on.
If $|c|<1$, this "motion" of the wave-number grid with respect to $k_c$ causes strong variations of the values of factors $f(k)$ in fidelity, which results in pronounced oscillations of fidelity, see Fig. \ref{Fid_mu0333_c05_d10_od_L}.
In an analogous situation encountered in a one-dimensional XY model,
similar oscillations have been ingeniously described by Rams and Damski \cite{rams PRA 11} in a quantitative  manner; they provided  accurate approximate formulae for fidelity of sufficiently large systems.
Therefore we do not dwell upon this point.

For $|c| \geq 1$, the oscillations of fidelity are much less pronounced, however for small $|\delta|$ and not too large systems they can still be significant. Fig. \ref{f_(0.5,0)_(0,1)} makes clear that for given $\delta$ the oscillations are damped with increasing system size $L$, which can be seen in  Fig. \ref{Fid_mu0333_c05_d10_od_L}.

\begin{figure}
\begin{center}
\includegraphics[width=8cm,clip=on]{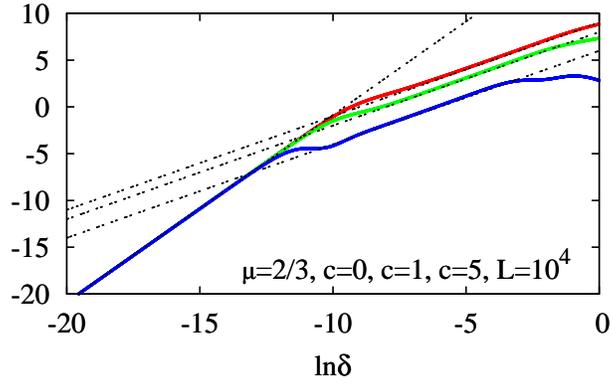}
\caption{\label{Fid_mu0666_d10_od_d} (Color online). The system size is $L=10^4$.
Plot of $-\ln {\mathscr{F}}_{(0,1)}((2/3,0), \delta)$ versus $\delta $ in doubly logarithmic scale, for three values of $c$; from bottom to top: $c=5$ -- blue line, $c=1$ -- green line, $c=0$ -- red line. The three parallel black dotted  straight lines of slope $1$ indicate the macroscopic-system region,
where $-\ln {\mathscr{F}}_{(0,1)}((2/3,0), \delta ) \sim \delta $. The black dotted straight line of slope $2$ represents formula (\ref{Fid_(m,0)_ss}) with adjusted numerical coefficient and indicates the small-system regime, where it coincides with the three plots. }
\end{center}
\end{figure}

As long as fidelity oscillates strongly with $L$, such notions as fidelity susceptibility and finite-size scaling have no meaning usually attributed to them. However, we can assign a meaning to those notions as follows.
Fidelity attains a local maximum if $k_c$ is located approximately in the middle between two consecutive points of the grid, that is the closest to $k_c$ wave numbers are approximately $k_c \pm \pi/L$. These wave numbers give dominant contributions to fidelity for sufficiently small $|\delta|$. Taylor expanding  $\ln f(k_c \pm \pi/L)$ in $\delta$ and summing contributions from those two wave numbers we obtain the following approximation to fidelity:
\begin{equation}
-\ln {\mathscr{F}}_{(0,1)}((\mu,0), \delta) \approx  \frac{2}{\pi^2}(\mu \delta L)^2.
\label{Fid_(m,0)_ss}
\end{equation}
The above formula provides an envelope of minima of $-\ln {\mathscr{F}}_{(0,1)}((\mu,0), \delta)$ as function of $L$. For given $\delta$ and a range of system sizes, that are not too  large and not too small (see our comment below), it fits very well numerical data, provided $\mu$ is sufficiently small, say $\mu < 0.1$; for larger $\mu$, the numerical coefficient $2/\pi^2$ has to be increased suitably, see Fig. \ref{Fid_mu0333_c05_d10_od_L}.
The set of $\delta$ and $L$, for which formula (\ref{Fid_(m,0)_ss}) approximates well the envelope of minima of
$ -\ln {\mathscr{F}}_{(0,1)}((\mu,0), \delta)$ as function of $L$, for fixed $\delta$, can be qualified as the small-system regime.
In Fig. \ref{Fid_mu0666_d10_od_d} we show that, in the small-system regime, formula (\ref{Fid_(m,0)_ss}) with adjusted numerical coefficient provides as well a good approximation to  $ -\ln {\mathscr{F}}_{(0,1)}((\mu,0)$ as function of $\delta$, for fixed $L$. Therefore, we can say, in the specified above sense, that in the small-system regime $-\ln {\mathscr{F}}_{(0,1)}((\mu,0), \delta)$ scales with $\delta$ and $L$ as $(\delta L)^2$.

In Fig. \ref{Fid_mu0333_c05_d10_od_L} (left panel) it is well seen that for fixed $\delta$ the range of $L$, for which we deal with a small system in the above sense, is bounded from below.
For smaller $L$, $-\ln {\mathscr{F}}_{(0,1)}((\mu,0), \delta)$ scales as $L$,  the system enters a new regime, which we call the ultra small-system regime. The crossover from the $L^2$ scaling to $L$ scaling  on decreasing sufficiently $L$, i.e. small-system-ultra small-system crossover, can be attributed to the proximity of the multicritical point.
This is demonstrated in Fig. \ref{Fid_mu_c05_d10_od_L} (right panel) when the critical point $(\mu,0)$ approaches the multicritical point $(0,0)$ the region of $L$ scaling expands, incorporating larger values of $L$. In the limit $\mu \to 0$ the small-system regime (where $L^2$ scaling holds) disappears, $\lim_{\mu \to 0} -\ln {\mathscr{F}}_{(0,1)}((\mu,0), \delta)$ is well approximated by formula (\ref{F_(0,J) (0,1)}) with $J=0$.
Fig. \ref{Fid_mu_c05_d10_od_L} (right panel) indicates also that the small-system-ultra small-system crossover, on decreasing sufficiently $L$, occurs in a vicinity of the intersection of the two black dotted lines, given analytically by formulae (\ref{F_(0,J) (0,1)})  with $J=0$ and (\ref{Fid_(m,0)_ss}); hence the considered crossover condition reads:
\begin{equation}
\mu^2 L \sim 1.
\label{L_2L_cross}
\end{equation}

Concerning the macroscopic-system regime, $-\ln {\mathscr{F}}_{(0,1)}((\mu,0), \delta)$  scales with $\delta$ and $L$ as $|\delta| L$, see Fig. \ref{Fid_mu0666_d10_od_d}. Finally, we note that the small- macroscopic-system crossover takes place if the expected crossover condition $L/{\tilde{\xi}}_{\bs e}({\bs\lambda}, \delta) \sim 1$ holds.

 Summing up, in the generalized sense, explained above, in transitions across $J {=} 0$ critical line, near ${\bs \lambda}_c {=} (\mu,0)$, scaling laws (\ref{susc scaling 3}) and (\ref{fid scal}) together with crossover condition (\ref{crossover}) set by $\xi$ (\ref{xi}) are obeyed with $\nu=1$. On decreasing sufficiently $L$, a crossover to a new regime, ultra small-system regime, takes  place.

%\clearpage

\subsection{\label{cep} Across the $J {=} 0$ critical line, at the critical end point: ${\bs \lambda}_c {=} (1,0)$, $\bs e {=} (0,1)$}

As mentioned in section \ref{model}, in this case the critical index $\nu=1/2$.
At the critical point $(1,0)$ the excitation gap $E_k$ closes at $k_c=0$. In a line neighborhood of this critical point, in direction $(0,1)$, the function $f(k)$ assumes the form
\begin{eqnarray}
f(k)=\frac{1}{2}\left \{ 1+\frac{(\cos k - 1)^2 + J\tilde{J} \cos^2k}
{\sqrt{ \left[ (\cos k - 1)^2+J^2\cos^2 k \right] \left[(\cos k - 1)^2 + \tilde{J}^2\cos^2 k\right]}}
\right \},
\label{(1,0)_f}
\end{eqnarray}
where we set
\begin{eqnarray}
J=(c-1)\delta, \quad \tilde{J}=(c+1)\delta.
\end{eqnarray}

\begin{figure*}\centering
\begin{subfigure}[b]{0.48\textwidth}\centering
\includegraphics[width=8cm,clip=on]{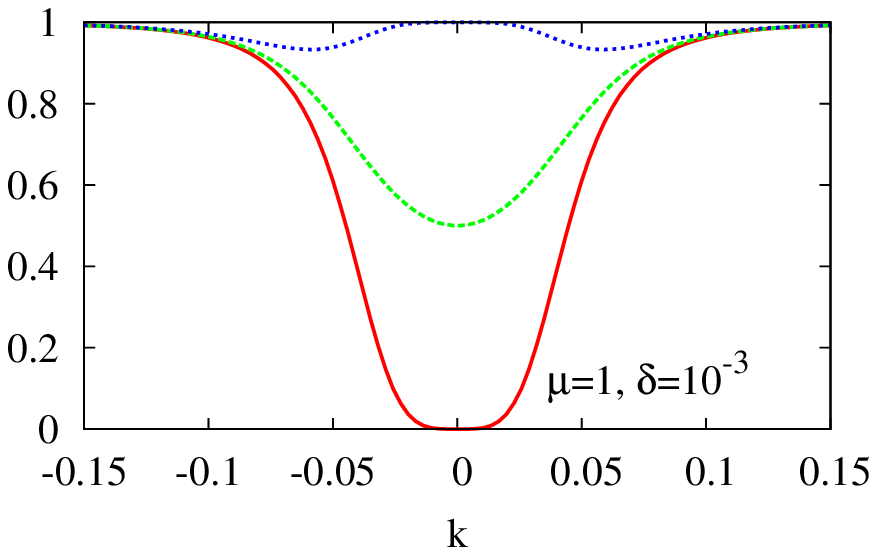}
\end{subfigure}
~
\begin{subfigure}[b]{0.48\textwidth}\centering
\includegraphics[width=8cm,clip=on]{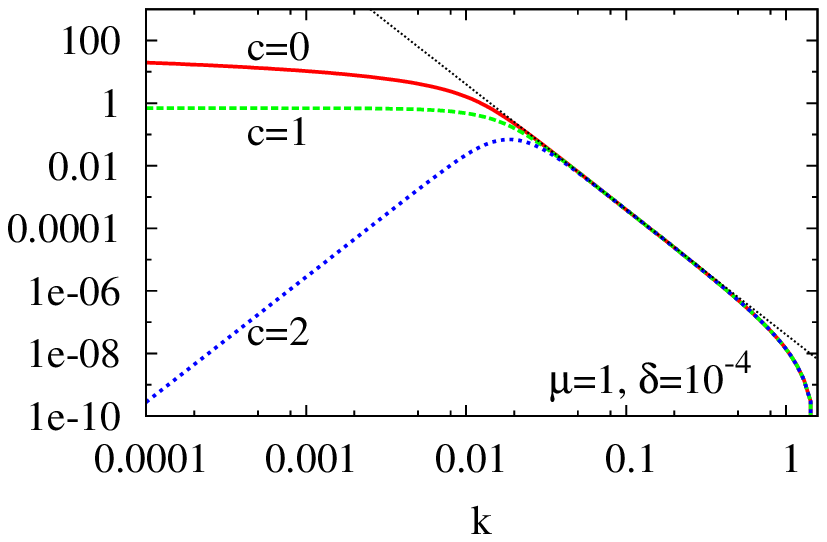}
\end{subfigure}
\caption{\label{f_(1,0)_(0,1)}\label{h_(1,0)_(0,1)} The plots of $f$ (left panel) and
$-\ln f$ (right panel) versus $k$ for three values of $c$; from bottom to top: $c=0$
-- red continuous line, $c=1$ -- green dashed line, $c=2$ -- blue dotted line.
Right panel is in doubly logarithmic scale and the black dotted line of slope $-4$
indicates the region of $-\ln f(k) \sim k^{-4}$ behavior.}
\end{figure*}

Apparently, $f(k)$ is continuous for all $c$, and away from $k=0$ it is insensitive to parameter $c$,
see Fig.~\ref{f_(1,0)_(0,1)}. Moreover, away from $k=0$ it behaves like $k^{-4}$ (right panel of Fig.~\ref{h_(1,0)_(0,1)}). Consequently, in the small-system regime $-\ln {\mathscr{F}}_{(0,1)}((1,0), \delta)$ does not depend on $c$ and scales with the system size as $L^4$. To get an explicit approximation to $-\ln {\mathscr{F}}_{(0,1)}((1,0), \delta)$ in the small-system regime, in the lowest order in $\delta$, it is enough to use the approximation $\cos k \approx 1 - k^2/2$, Taylor expand $f(k)$ in $\delta$ at $\delta=0$, and take into account the contributions of the two closest to $k_c=0$ wave numbers $\pm \pi/L$:
\begin{eqnarray}
-\ln {\mathscr{F}}_{(0,1)}((1,0), \delta) \approx \frac{4}{\pi^4} \delta^2 L^4 .
\label{(1,0)_Fid_ss}
\end{eqnarray}
In the macroscopic-system regime, the integral (\ref{fid_integral}) can be approximated by
$|\delta|^{1/2} {\cal{A}}(c)$, that is
\begin{eqnarray}
-\ln {\mathscr{F}}_{(0,1)}((1,0), \delta) \approx  |\delta|^{1/2} L {\cal{A}}(c),
\label{(1,0)_Fid_ms}
\end{eqnarray}
where the universal function ${\cal{A}}(c)$ is given by
\begin{eqnarray}
{\cal{A}}(c) = \frac{2}{\pi} \int_0^{\infty} dx \frac{x^4}{x^8+8c^2x^4+16(c^2-1)^2} =
\frac{(c^2+\sqrt{2c^2-1})^{1/4} - (c^2-\sqrt{2c^2-1})^{1/4}}{8\sqrt{2c^2-1}}.
\label{(1,0)_A)}
\end{eqnarray}
Formula (\ref{(1,0)_A)}) holds for all $c$. The function ${\cal{A}}(c)$ is positive everywhere, at $c=0$ it amounts to $\cos(3\pi/8)/4$, and at $c=1$ its value is $2^{-11/4}$. We note that for $c=1$, that is when one of the ground states used for calculating fidelity is the ground state at the critical point, ${\cal{A}}(c)$ is nondifferentiable.

 In Fig. \ref{Fid_(1,0)_(0,1)} we show that numerical data compare well with the approximations given by formulae (\ref{(1,0)_Fid_ss}) and (\ref{(1,0)_Fid_ms}).
 Thus, in transitions across the $J {=} 0$ critical line, near the critical end point scaling laws (\ref{susc scaling 3}) and (\ref{fid scal}) together with crossover condition (\ref{crossover}) set by $\xi$ (\ref{xi}) are obeyed with $\nu=1/2$.

\begin{figure}
\begin{center}
\includegraphics[width=8cm,clip=on]{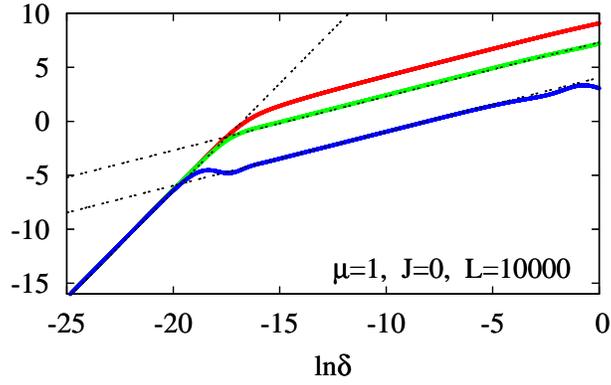}
\caption{\label{Fid_(1,0)_(0,1)}  The system size is $L=10^4$. The plot of $-\ln{\mathscr{F}}_{(0,1)}((1,0)_c, \delta))$  versus $\delta$ for three values of $c$;  from bottom to top: $c=2$ -- blue dotted line, $c=1$ -- green dashed line, $c=0$ -- red continuous line. The black dotted straight line of slope $2$ represents the small-system analytic approximation (\ref{(1,0)_Fid_ss}). The two parallel  black dotted straight lines of slope $1/2$ represent the macroscopic-system analytic approximation (\ref{(1,0)_Fid_ms}).
Scaling is consistent with $\nu=1/2$,
i.e. $-\ln {\mathscr{F}}_{(0,1)}((1,0), \delta)\sim\delta^2 L^4$ in small-system regime and
$-\ln {\mathscr{F}}_{(0,1)}((1,0), \delta)\sim |\delta|^{1/2}L$ in macroscopic-system regime.}
\end{center}
\end{figure}

\section{\label{summ} Summary}

In this paper we have presented results of detailed studies of critical behavior of quantum fidelity in a BCS-like model. We have limited our presentation to a one-dimensional version of this model, where we have been able to derive a number of analytical results for correlation functions and quantum fidelity in critical regions. Needless to say that analytical results facilitate a lot understanding of numerical data. The presented results, obtained for a model that is out of the realm of paradigmatic  quantum spin chains, can be thought of as complementary to those obtained by Rams and Damski \cite{rams PRL 11}, \cite{rams PRA 11}. Results pertaining to the two-dimensional case, where almost no aid of analytical results is available, but still the exact diagonalizability of the model plays a key role, are in preparation and will be published elsewhere.

\begin{table}
\begin{tabular}{l|c|c|c|c}
Critical region & Small s.& Macroscopic s.& Crossover & $\nu$ \\
\hline
${\bs \lambda}_c=(0,J)$, ${\bs e}=(1,0)$, large $|J|$ & $(\delta L)^2$ & $|\delta|L$ & $L \sim \xi^{(1)}$ & $1$\\
${\bs \lambda}_c=(0,J)$, ${\bs e}=(1,0)$, small $|J|$ & $(\delta L)^2$ & $|\delta|L$ & $L \sim \xi'^{(2)}$ & $1$\\
${\bs \lambda}_c=(0,J)$, ${\bs e}=(0,1)$ & $\delta^2 L$ & $\delta^2 L$ & no crossover & undef.\\
${\bs \lambda}_c=(0,0)$, various ${\bs e}\neq(0,1)$ & $\delta^2 L$ & $|\delta|L+|\delta|\ln|\delta| L$ & $L \sim \pi/\delta\cos\phi$ & undef.\\
${\bs \lambda}_c=(\mu, 0)$, ${\bs e}=(0,1)$,  large $|\mu|^{(3)}$ & $(\delta L)^2$ & $|\delta|L$ & $L \sim \xi$ & $1$\\
${\bs \lambda}_c=(1, 0)$, ${\bs e}=(0,1)$ & $\delta^2L^4$ & $|\delta|^{1/2}L$ & $L \sim \xi$ & $1/2$
\end{tabular}
\caption{\label{resume}Resume of $-\ln {\mathscr{F}}_{\bs e}({\bs\lambda}_c, \delta)$ scaling in small-system and
macroscopic-system regimes, for critical regions investigated in the paper, except the case of
${\bs \lambda}_c=(\mu, 0)$ being near the multicritical point (see section \ref{im}).
$^{(1)}$ The correlation length $\xi$ is specified in sections \ref{model} and \ref{lJ}.
$^{(2)}$ The length $\xi'$ is defined in section \ref{sJ}.
$^{(3)}$ This scaling holds in the generalized sense (see section \ref{im}).}
\end{table}

We have verified critical scaling laws of quantum fidelity in the whole range of critical behavior: the critical scaling law (\ref{susc scaling 2}) in the small-system regime, where the effective linear size of the system satisfies the condition $L/{\tilde{\xi}}_{\bs e}({\bs\lambda}, \delta) \ll 1$, then the crossover condition $L/{\tilde{\xi}}_{\bs e}({\bs\lambda}, \delta) \sim 1$, and finally the critical scaling law (\ref{fid scal}) in the macroscopic-system regime, where $L/{\tilde{\xi}}_{\bs e}({\bs\lambda}, \delta) \gg 1$. In the latter case, performing the thermodynamic limit, $L \to \infty$ for fixed $\delta$,  we have found that, for all the considered critical points, fidelity vanishes as $-\ln {\mathscr{F}}_{\bs e}({\bs\lambda}_c, \delta) \sim  L$, that is the involved ground states become orthogonal; this is known as the Anderson orthogonality catastrophe \cite{anderson 67} and has interesting consequences for various condensed-matter systems.

First, in transitions across the $J$-axis ($\nu=1$ -- exact result), sections \ref{lJ} and \ref{sJ}, and across the $\mu$-axis but for the critical end point (${\bs \lambda}_c=(1,0)$, $\nu=1/2$ -- numerical result), section \ref{cep}, fidelity scales according to the laws (\ref{susc scaling 2}) and (\ref{fid scal}), in the small-system and macroscopic-system regimes, respectively. The behavior of fidelity in these  transitions is fairly analogous to that in transitions across the line of critical points with specific value of transverse magnetic field ($g=1$) in anisotropic XY chain \cite{rams PRA 11}. The crossover condition (\ref{crossover}) with the correlation length $\xi$ given by (\ref{xi}) is satisfied provided  that the underlying critical point is not in a close proximity with the multicritical point. In the opposite case a characteristic length $\xi'$, different from the correlation length $\xi$, determines the crossover condition.

Second, we encounter a peculiar critical fidelity scaling in transitions that occur in neighborhoods of the multicritical point ${\bs \lambda}_c=(0,0)$, section \ref{mcp}, which is neither (\ref{susc scaling 2}) nor (\ref{fid scal}). Specifically, in the small-system regime the anomalous scaling
$-\ln {\mathscr{F}}_{\bs e}({\bs\lambda}_c, \delta) \sim  \delta^2 L$ holds. The crossover condition is consistent with $\nu=1$.
In the macroscopic-system regime,
$-\ln {\mathscr{F}}_{\bs e}({\bs\lambda}_c, \delta)$ is nonanalytic in $\delta$ and the dependence on $\delta$ is even not power-like. This behavior differs considerably from that in anisotropic XY chain \cite{rams PRA 11}.

Third, an anomalous fidelity behavior takes place also in transitions along the $J$-axis, section \ref{Jaxis},  which is the line of critical points. Independently of the location, with respect to the multicritical point, of the critical point on this line, the anomalous scaling, found in the small-system regime in neigborhoods of the multicritical point,
$-\ln {\mathscr{F}}_{\bs e}({\bs\lambda}_c, \delta) \sim  \delta^2 L$, holds with no small- macroscopic-system crossover. The latter property is consistent with the correlation length being infinite. Such an anomalous scaling has been found also at the line $g=1$ of critical points in anisotropic XY chain \cite{rams PRA 11}, but only away from the multicritical point.

Fourth, in transitions across the $\mu$-axis ($\nu=1$ -- numerical result), section \ref{im}, fidelity as function of $L$ may exhibit large-amplitude oscillations in the small-system regime, which makes such notions as fidelity susceptibility and finite-size scaling ill defined. One can assign a meaning to them in the sense of the envelope of minima of $-\ln {\mathscr{F}}_{\bs e}({\bs\lambda}_c, \delta)$ as function of $L$, and then, in the new sense, the scaling law (\ref{susc scaling 2}) is obeyed. The crossover condition (\ref{crossover}) and macroscopic-system scaling (\ref{fid scal}) hold  as well but in the new sense. Interestingly, an influence of the multicritical point can be seen when the considered  critical point approaches the multicritical one (i.e. $\mu$ approaches zero), with $L$ and $\delta$ fixed and  sufficiently small. Then, the small-system $L^2$-scaling (\ref{susc scaling 3})  changes to $L$-scaling, as it is in a neighborhood of the multicritical point,  which we call the small-system - ultra small-system crossover; the crossover condition reads $\mu^2 L \sim 1$.

 Finally, for reader's convenience we present in Table~\ref{resume} a resume of our results concerning critical scaling of $-\ln {\mathscr{F}}_{\bs e}({\bs\lambda}_c, \delta)$ in small-system and macroscopic-system regimes, in various critical regions investigated in the paper.

\begin{center}
{\bf Acknowledgements}\\
The presented studies have been supported by the University of Wroclaw through the project Nr 2288/M/IFT/12.
\end{center}

\end{document}